\documentclass[12pt, a4paper]{article}
\pdfoutput=1
\usepackage{pstricks}
\usepackage{cite}
\usepackage{hyperref}
\usepackage{ dsfont }
\usepackage{amsmath}
\usepackage{amsfonts}
\usepackage{amssymb}
\usepackage{latexsym}
\usepackage{graphicx}
\usepackage{color}
\usepackage{enumerate}
\usepackage{multirow}
\usepackage{colortbl}
\usepackage{url}
\usepackage{enumitem}
\usepackage{comment}
\usepackage{tabu}
\usepackage{array}
\usepackage{subfigure}

\def\be   {\begin{equation}}   \def\ee   {\end{equation}}
\def\ba   {\begin{array}}      \def\ea   {\end{array}}
\def\bea  {\begin{eqnarray}}   \def\eea  {\end{eqnarray}}
\def\bean {\begin{eqnarray*}}  \def\eean {\end{eqnarray*}}

\def\to {\rightarrow}

\newcommand{\gev}{\ensuremath{\,\mathrm{GeV}}}

\newcommand{\kpc}{\ensuremath{\,\mathrm{kpc}}}

\newcommand*{\mkblue}[1]{{\color{blue}{#1}}}

\newcommand*{\mkpur}[1]{{\color{purple}{#1}}}

\numberwithin{equation}{section}

\begin{document}

\vspace{0cm}
%\hfill IPMU14-0337 \\

\begin{center}
\vspace{2cm}

{\Large
Inverting cosmic ray propagation by Convolutional Neural Networks    
}
\\ [2.5cm]
{\normalsize{\textsc{
Yue-Lin Sming Tsai$^{\,a,}$\footnote{\textsl{smingtsai@pmo.ac.cn}}, 
Yi-Lun Chung$^{\,b,}$\footnote{\textsl{s107022801@m107.nthu.edu.tw}}, 
Qiang Yuan$^{\,a,c,d,}$\footnote{\textsl{yuanq@pmo.ac.cn}},\\
Kingman Cheung$^{\,b,e,}$\footnote{\textsl{cheung@phys.nthu.edu.tw}}
}}}
\\[1cm]

\normalsize{\textit{
$^{a}$~Key Laboratory of Dark Matter and Space Astronomy,  
Purple Mountain Observatory, Chinese Academy of Sciences, Nanjing 210023, China\\ \vspace{1.5mm}
$^b$ Department of Physics, National Tsing Hua University,
Hsinchu 300, Taiwan \\ \vspace{1.5mm}
$^c$ School of Astronomy and Space Science, University of Science and
Technology of China, Hefei 230026, China \\ \vspace{1.5mm}
$^d$ Center for High Energy Physics, Peking University, Beijing 100871, China \\ \vspace{1.5mm}
$^e$ Division of Quantum Phases and Devices, School of Physics, 
Konkuk University, Seoul 143-701, Republic of Korea \\ \vspace{1.5mm}
}}
{ \large{\textrm{
Abstract
}}}
\\ [0.5cm]
\end{center}

We propose a machine learning method to investigate the propagation of cosmic 
rays based on the precisely measured spectra of the primary and secondary cosmic 
ray nuclei of Li, Be, B, C, and O from AMS-02, ACE, and Voyager-1. We train two 
convolutional neural networks. One network learns how to infer propagation and 
source parameters from the energy spectra of cosmic rays, and the other network, 
which is similar to the former, has the flexibility to learn from the data with 
added artificial fluctuations. Together with the simulated data generated by 
\texttt{GALPROP}, we find that both networks can properly invert the propagation 
process and infer the propagation and source parameters reasonably well.
This approach can be much more efficient than the traditional Markov chain
Monte Carlo fitting method for deriving the propagation parameters if users 
choose to update confidence intervals with new experimental data.
%Our publicly available network suggests a narrower prior distribution of CR model 
%parameters based on the proposed likelihood distribution of the measured spectra
%as long as each new error bar is not more than $\sqrt{10}$ times larger than the 
%present one. In addition, each new central value must only fluctuate within 
%$\pm \sqrt{10}$ times the present error bar. If the two above validity conditions 
%break, the network may extrapolate an untested result.
Both of the trained networks are available at  (\href{https://github.com/alan200276/CR_ML}{\url{https://github.com/alan200276/CR_ML}}).

\def\thefootnote{\arabic{footnote}}
\setcounter{footnote}{0}
\pagestyle{empty}

\newpage
\pagestyle{plain}
\setcounter{page}{1}

%%%%%%%%%%%%%%%%%%%%%%%%%%%%%%%%%%%%%
\section{Introduction}\label{sec:intro}
%%%%%%%%%%%%%%%%%%%%%%%%%%%%%%%%%%%%%

The source spectrum and propagation mechanism of cosmic rays (CRs) remain largely unknown.
With the exception of the Voyager spacecraft, which may have travelled to the outside of the
heliosphere \cite{Stone:2013}, all other CR data are measured within Earth's neighbourhood.
Based on observational data, the source and propagation parameters of CRs can degenerate each other,
which barely renders a direct measure of such parameters.
The traditional parameter estimation procedure involves first building a propagation model of CRs, parameterizing
the source spectra and propagation processes, and then fitting the data (e.g., \cite{Maurin:2001sj,Trotta:2010mx,Yuan:2017ozr}).
However, traditional procedures are time-consuming from a computational perspective, especially when
numerical simulations of CR propagation using simulators such as
\texttt{GALPROP}\footnote{\url{http://galprop.stanford.edu/}} \cite{Strong:1998pw}
or \texttt{DRAGON}\footnote{\url{http://www.dragonproject.org/Home.html}}~\cite{Evoli:2008dv}
is employed. 
Furthermore, in relevant studies with similar issues, 
such as the search for dark matter annihilation or decay signals, 
the uncertainty computation of the background models that are often ignored, 
and only some benchmark settings are investigated
due to the difficulties in obtaining full uncertainty bands
(see \cite{Cui:2016ppb,Cuoco:2016eej}, who attempted studies to include the uncertainties described above).
Increasing the speed of a large number of repetitive computations is a great challenge
for CR physics studies and related particle physics problems.

The general idea of global fitting is to
project the probability distribution of the measured CR fluxes
to the model parameter space.\footnote{
The probability distribution can be described by the chi-square $\chi^2$ or likelihood
statistics for the Frequentist interpretation, while the Bayesian interpretation is based on the posterior distribution. 
Note that the Gaussian likelihood probability density
can be written as $\sim \exp\left(-\frac{\chi^2}{2}\right)$.}
One can use various sampling tools, such as the
Markov chain Monte Carlo (MCMC) method or nest sampling, to explore the parameter space.
The chi-square $\chi^2$ statistic is built based on fluxes instead of the model parameter space. Hence, too many samples may be used in a trial and error approach in an unwanted region.
Moreover, once the likelihood distribution is changed,
it is better to restart the whole analysis to obtain good coverage.
If there are only mild changes in the likelihood distribution,
there are several approaches to efficiently and quickly complete new scans.
For example, the old posterior distribution can be used as an updated prior distribution
for the new scans or the covariance matrix generated by old scans
can be adopted as the initial prior distribution.
However, these approaches still require performing the parameter scan again.

To avoid these additional scanning procedures,
our goal is to obtain the corresponding propagation and
source parameters directly by using spectra of interest from experimental data.
The likelihood distribution in CR spectra can be directly taken
from experimental measurements without exploring the CR model parameter space.
Therefore, we can use the inverse function of the CR propagation equation to calculate the corresponding propagation and
source parameters directly from the desired CR spectra.
Nevertheless, CR propagation is a complicated partial differential equation, and
it is not easy to obtain its inverse function.
In this work, we show that machine learning (ML) is a powerful technique that can be used to undertake this task by
interpolating a large data set, especially in a multidimensional parameter space.

We can replace the propagation simulations with the ML interpolation function 
as long as a massive dataset is produced before learning~\cite{Lin:2019ljc}.
Therefore, we expect the learning networks to address the time-consuming tasks of global fitting and generating massive simulations.
Several groups have made strong efforts in searching for new physics beyond the Standard Model, see for
example, Refs.~\cite{Ren:2017ymm,Abdughani:2019wuv,Alsing:2019xrx,Lei:2020ucb}.
The purpose of this work is to convert experimental data to model parameter distributions directly using ML techniques.

In recent decades, ML methods such as support vector machines~\cite{BOSER,VAPNIK}
and boosted decision trees (BDTs)~\cite{Friedman,Ridgeway,Chen_2016} show good performance
in classification and regression problems.
In high-energy particle experiments, BDTs are more popular 
than support vector models because they are an interpretable model.
The performance of BDTs can be examined from input variable distributions such as invariant mass and
transverse momentum\cite{Guest:2018yhq}.
More recently, neural network learning has become a leading tool in every field
because a neural network can provide a more flexible variety of architectures to obtain hidden information from data.
Dense neural networks (DNNs) use several layers consisting of many neurons to learn the principle components in the data.
Convolutional neural networks (CNNs) are better than DNNs 
in handling the correlated information of pixelwise or multiple one-dimensional data.
In addition, graph neural networks~\cite{DBLP:journals/corr/abs-1812-08434} can build classification rules
or regression predictions from point data without a fixed shape.
Furthermore, recurrent neural networks and long short-term memory networks~\cite{DBLP:journals/corr/abs-1808-03314} aim
to cope with time sequence problems.
A recently developed architecture called Transformer~\cite{Polosukhin},
shows powerful performance in many problems, for instance, the precise identification of top jets~\cite{Fenton:2020woz}.
In this work, we adopt a CNN architecture to invert the CR propagation by taking
CR spectra and model parameters as inputs and outputs of the network,
respectively. A CNN uses a convolution operation and multiple networks with a non-polynomial
activation function and is thus a well-designed tool to address not only the
correlations between CR spectra (different elements and different energy bins)
but also the high-dimensional inverse-propagation problem.

In this work, we build a CNN network to determine a multidimensional inverse function
of CR propagation. We first generate simulated data for ML using MCMC scans~\cite{Yuan:2017ozr,Yuan:2018vgk}.
To demonstrate the capability of updatable measurements,
we enlarge the uncertainties of the current CR data by a factor of $\sqrt{10}$ in
the MCMC scans. 
To a certain degree, we are deprived of understanding with respect to new data.
We assume that the parameter space allowed by the current CR data,
with such enlarged uncertainties, should cover the new measurements, which have smaller error bars.

After building the proposed network, the inverse function of CR propagation (CNN version) allows us
to quickly guess the experimental data distribution in the CR model parameter space without performing a new MCMC scan.
As a comparison, we introduce two networks: one network learns from
the spectra directly, while the other network learns from this data but with some artificial fluctuations.
These networks can predict the allowed model parameter space
using actual experimental data (with original uncertainties).

The primary goal of this work is to build an inverse function using the machine learning approach, 
such that the inverse function can readily invert observational datasets into the corresponding model parameter space.
The advantage of this method is emphasized as follows.
{\it When a new dataset is available or an existing dataset is updated,
the proposed networks do not need to be retrained and the whole analysis does not need to be repeated.
The trained machines can easily guess the high probability region of the parameter space.}

This paper is organized as follows.
In Sec.~\ref{sec:propagation}, we review the framework of CR propagation.
The model parameters used in this work are defined and explained.
Next, we briefly introduce the CNN in Sec.~\ref{sec:ML}.
Subsequently, we show the architecture of the two CNNs in Sec.~\ref{sec:method}.
The generation of mock data for training and pseudo-data for testing are described in detail.
Then, we show the usage of the networks to invert the propagation of CRs.
In Sec.~\ref{sec:result}, the performance of the two networks and a comparison
with the traditional global fitting method are presented.
Finally, we summarize our work in Sec.~\ref{sec:conclusion}.

%%%%%%%%%%%%%%%%%%%%%%%%%%%%%%%%%%%%%
\section{Cosmic Ray propagation}\label{sec:propagation}
%%%%%%%%%%%%%%%%%%%%%%%%%%%%%%%%%%%%%

Charged CRs diffuse in the random magnetic field of the Milky Way
and interact with the interstellar medium (ISM). During propagation,
CR particles acquire or lose energy and fragment into secondary particles
due to interactions with the ISM and magnetic fields. The general form of the
diffusion equation of CRs in the Milky Way with the source function
$Q({\bf x},p)$ is~\cite{Strong:2007nh}
\begin{eqnarray}
\frac{\partial \psi}{\partial t}
&=& Q({\bf x},p)+\nabla\cdot(D_{xx}\nabla \psi-{\bf
V_c}\psi)+\frac{\partial}{\partial p}p^2D_{pp}\frac{\partial}
{\partial p}\frac{1}{p^2}\psi \nonumber \\
&-& \frac{\partial}{\partial p}
\left[\dot{p}\psi-\frac{p}{3}(\nabla\cdot{\bf V_c}\psi)\right]-
\frac{\psi}{\tau_f}-\frac{\psi}{\tau_r}, \label{eq:prop}
\end{eqnarray}
where $\psi$ is the CR differential number density per unit momentum interval,
${\bf V_c}$ is the convection velocity$, \dot{p}\equiv{\rm d}p/{\rm d}t$
is the momentum loss rate, and $\tau_f$ and $\tau_r$ are the time scales
for fragmentation and radioactive decay, respectively.
Considering a particle with momentum $p$ and charge $Ze$, we assume a
spatially homogeneous diffusion coefficient $D_{xx}$ as~\cite{Maurin:2010zp}
\begin{equation}
D_{xx} = D_0\beta^\eta \left( R/R_0 \right)^{\delta}, 
\end{equation}
where $\beta\equiv v/c$ is the particle velocity at light speed $c$,
$R\equiv pc/Ze$ is the rigidity, $\delta$ describes the rigidity-dependence
of the diffusion coefficient, and $D_0$ is a normalization parameter.
The parameter $\eta$ gives a phenomenological modification of the diffusion
coefficient at low energies to better fit the data~\cite{DiBernardo:2009ku}.
The momentum diffusion term describes the reacceleration of CRs during propagation, 
where the diffusion coefficient $D_{pp}$ relating to $D_{xx}$ as~\cite{Seo:1994}
\begin{equation}
D_{pp}D_{xx}=\frac{4p^2v_A^2}{3\delta(4-\delta^2)(4-\delta)},
\end{equation}
where $v_A$ is the Alfven speed.
We assume that CRs are confined within a cylinder with a radius of $R_h=20$~kpc and a thickness of $2z_h$.
CRs outside this region are assumed to escape freely.
Ref.~\cite{Yuan:2018lmc} showed that the reacceleration model can fit the currently available data well.
However, propagation with significant convection is disfavoured.
Therefore, we restrict the current work to the reacceleration model configuration
and the main propagation parameters are $D_0$, $\delta$, $\eta$, $v_A$, and $z_h$.

We can parameterize the rigidity spectrum at the source as
\begin{eqnarray}
\label{eq:source2}
q_s=q_0\left(\frac{R}{R_{\rm br,1}}\right)^{-\nu_1}\left[1+\left(
\frac{R}{R_{\rm br,1}}\right)^2\right]^{(\nu_1-\nu_2)/2}
\left[1+\left(\frac{R}{R_{\rm br,2}}\right)^2\right]^{(\nu_2-\nu_3)/2}.
\end{eqnarray}
The parameters $q_0$, $\nu_1$, $\nu_2$, $\nu_3$, $R_{\rm br,1}$, and $R_{\rm br,2}$
are also treated as free parameters in this work.

Secondary-to-primary ratios, such as B/C and (Sc+Ti+V)/Fe, and the
unstable-to-stable ratios of secondary particles,
such as $^{10}$Be/$^9$Be and $^{26}$Al/$^{27}$Al are often used
to determine the propagation parameters~\cite{Strong:1998pw,Maurin:2001sj,Putze:2010zn}.
In this work, we use the Li, Be, B, C, and O fluxes
measured by AMS-02~\cite{Aguilar:2017hno,Aguilar:2018njt}, ACE~\cite{ACE,Zhu:2018jbk},
and Voyager~\cite{Cummings:2016pdr} to obtain the MCMC samples used for
training of the networks. We use the demodulated fluxes given in
Ref.~\cite{Zhu:2018jbk} to avoid addressing the solar modulation
effect\footnote{This is not crucial in this methodology-based paper.}.
In Ref.~\cite{Zhu:2018jbk}, a spline interpolation method was
employed to parameterize the wide-band spectra of various species
and a fit to the Voyager, ACE, and AMS-02 data was performed assuming
a force-field solar modulation model. Then, the data were demodulated
based on the best-fit modulation parameter and its associated uncertainty.
As described in the introduction, the error bars of these data have
been artificially enlarged by a factor of $\sqrt{10}$
to have a wide enough coverage of the parameter space.
For C and O, we assume that they share identical primary source
spectra, with different normalizations $q_0^{\rm C}$ and $q_0^{\rm O}$, respectively.
The propagation parameters determine the absolute fluxes of secondary nuclei.
However, as found in Ref.~\cite{Yuan:2018lmc}, the flux normalizations among different species may differ
by up to $20\%$ due to the uncertainties of the fragmentation cross-sections.
Thus, two additional constant factors $\xi_{\rm Li}$ and $\xi_{\rm Be}$ are further employed.
The total parameter space is 14 dimensional, including five propagation parameters,
five source spectral shape parameters, and four normalization parameters.
An example input file of the GALPROP simulator, the galdef file,
is available on the GitHub website\footnote{\url{https://github.com/alan200276/CR-ML/blob/main/MC_setup/galdef_54_crml}}.

%%%%%%%%%%%%%%%%%%%%%%%%%%%%%%%%%%%%%
\section{Convolutional neural network}\label{sec:ML}
%%%%%%%%%%%%%%%%%%%%%%%%%%%%%%%%%%%%%

To date, the CNN is one of the most widely used deep learning algorithms. Deep learning is a type of ML method.
Traditionally, global fittings are employed to scrutinize physical models
from observational data.
Then, we can learn the correlations between model parameters by
analysing the massive datasets generated by global fittings.
In contrast, the CNN is a modern technique that can perform this task better than humans
because it can mimic the properties of a physical model by training network with massive data. 
As long as one can provide enough information with well-defined input and output parameters for training,
the network can find underlying patterns or correlations 
based on a set of input parameters and then predict a set of output parameters.

There are three common usages of CNNs: interpolation, fast computation, and inverse functions.
The CNN is a good tool for interpolation in a hyperspace of multiple parameters.
When the network is trained even with discretized data,
the corresponding outputs can be obtained by providing any input parameters.
Together with the interpolation function, a trained
network can provide predictions very quickly.
However, ``fast" computations are based
on many ``slow" calculations to provide data for training.
CNN predictions are always faster than the actual computations starting from scratch unless recycling old data is nonbeneficial.
The CNN can invert a complicated partial differential equation,
especially in a high-dimensional parameter space.
Once the CNN obtains the inverse function,
the experimental likelihood distribution of the physical observables (cosmic ray fluxes)
can be translated to model parameters (propagation and source parameters).

The CNN is also capable of distinguishing the features of images,
and fits our purpose of determining the correlation between the different energy spectra.
Essentially, the CNN structure connects input and output layers with three main layers:
a \textit{convolution layer}, \textit{pooling layer}, and \textit{fully connected layer}.
These three layers play different roles in quickly capturing the features of the data.
We review their functions below.

The convolution layer converts the cosmic ray energy spectrum (an input tensor) into several feature maps via a feature detector (filter).
These feature maps form a new tensor with one additional index.
Since this new tensor contains more neurons than the input layer, it requires much more computation time and memory.
We need a pooling layer after the convolution layer to reduce the tensor dimensions.
Moreover, we use max pooling~\cite{maxpooling}, which only takes the maximum value from one kernel.
Several convolution and pooling layers can be utilized to achieve better performance.
Finally, a fully connected layer is used to flatten the tensors and return an array of outputs.

\section{Methodology}\label{sec:method}

\begin{figure}
\begin{center}
\includegraphics[width=0.90\textwidth]{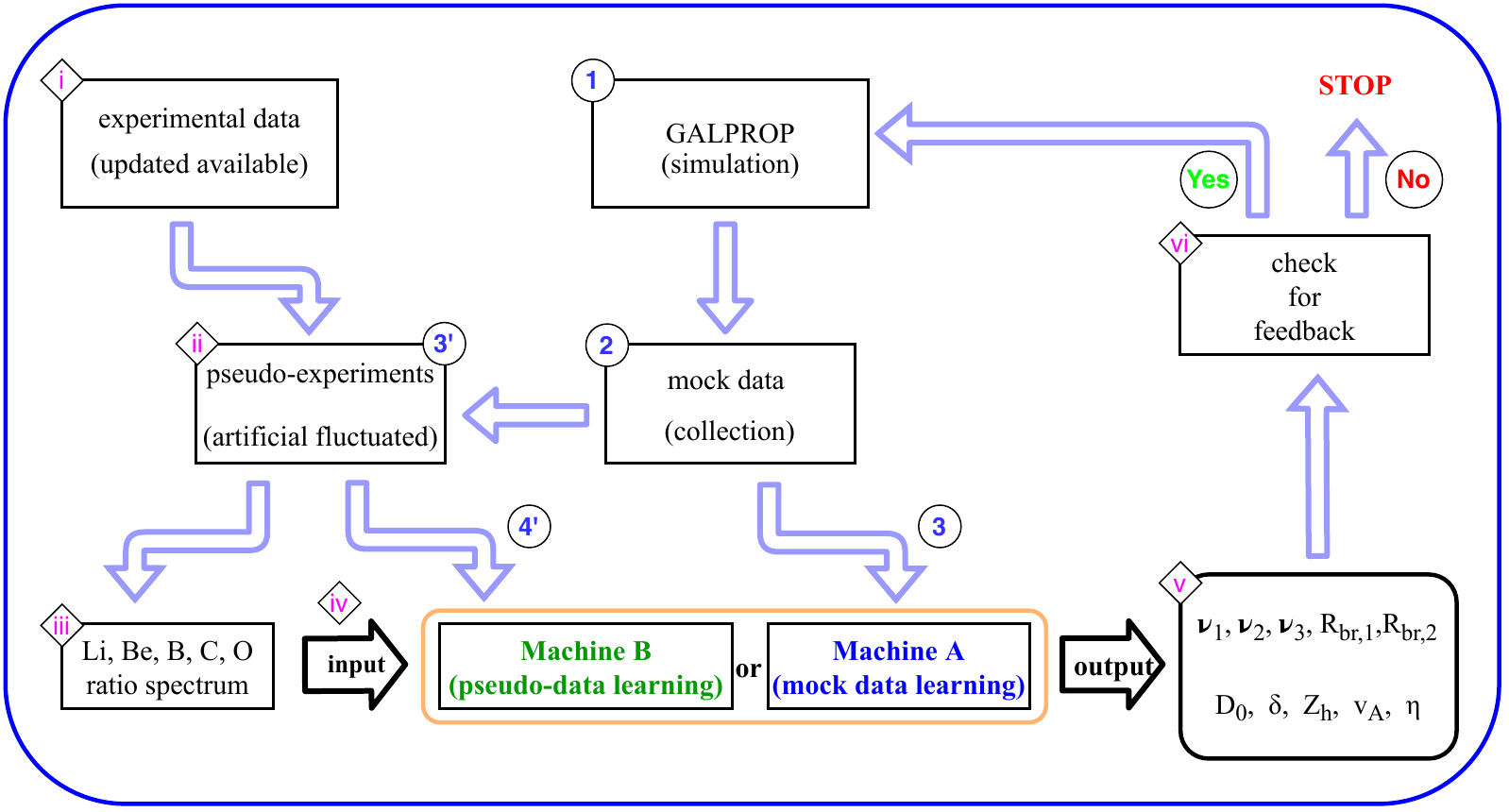}
\caption{Flowchart of our work. The mock/simulated data are generated using \texttt{GALPROP} 
and used to train two machines. The first machine is trained with the mock data directly.
The second machine is trained with pseudo-data (mock data with artificial fluctuations added).
The training procedure is $\mkblue{(1)}\to \mkblue{(2)}\to \mkblue{(3)}\to\mkpur{(v)}$ for {\bf Machine A}
and $\mkblue{(1)}\to \mkblue{(2)}\to \mkblue{(3^\prime)}\to \mkblue{(4^\prime)}\to\mkpur{(v)}$ for {\bf Machine B}.
The usage of two machines is $\mkpur{(i)}\to \mkpur{(ii)}\to \mkpur{(iii)}\to \mkpur{(iv)}\to \mkpur{(v)}$.
Finally, the feedback loop for modifying {\bf Machine A} is
$\mkpur{(i)}\to \mkpur{(ii)}\to \mkpur{(iii)}\to \mkpur{(iv)}\to \mkpur{(v)}\to \mkpur{(vi)}
\to \mkblue{(1)}\to \mkblue{(2)}\to \mkblue{(3)}\to\mkpur{(v)}\to\mkpur{(vi)}$.
For the feedback loop of {\bf Machine B}, we can replace
$\mkblue{(3)}$ with $\mkblue{(3^\prime)}\to \mkblue{(4^\prime)}$ in the feedback loop of {\bf Machine A}. 
\label{Fig:FlowChart}}
\end{center}
\end{figure}

The purpose of this work is to build a machine that can invert CR
propagation datasets into the allowed model parameters.
The major blocks of actions and data processing are depicted in Fig.~\ref{Fig:FlowChart}.
We will describe the ``\textit{mock data}" in Sec.~\ref{sec:fitting},
``\textit{pseudo-data}" in Sec.~\ref{sec:pseudoexp} and ~\ref{sec:preparation},
machine training in Sec.~\ref{sec:machines},
feedback data in Sec.~\ref{sec:feedback},
and data processing in Sec.~\ref{sec:processing}.

\subsection{Mock data}\label{sec:fitting}

We first generate a massive dataset for machine training to build the correlation  
between the energy spectra of CRs and model parameters. 
We use the \texttt{GALPROP} code to simulate the propagation, 
together with an MCMC parameter scan~\cite{Liu:2011re}. 

\begin{table}
\begin{center}
\begin{tabular}{|c|c|}
\hline
\hline
\multicolumn{2}{|c|}{Propagation parameters} \\
\hline\hline
Diffusion coefficient &  $1.0<D_{0}/10^{28}~{\rm cm^2~s^{-1}}<15.0$   \\
Diffusion coefficient rigidity power &  $0.2<\delta<1.0$   \\
Diffusion coefficient velocity power &  $-4.0<\eta<4.0$   \\
Alfven speed &  $0.0<v_A<60.0$   \\
Height of diffusion zone &  $1.0<z_h/{\rm kpc}<20.0$   \\
\hline\hline
\multicolumn{2}{|c|}{Source parameters} \\
\hline\hline
$1^{st}$ power-law index  & $0.0<\nu_1<2.0$      \\
$2^{nd}$ power-law index  & $1.0<\nu_2<4.0$      \\
$3^{rd}$ power-law index  & $1.0<\nu_3<4.0$      \\
First break break   & $2.0<\log (R_{\rm br,1}/{\rm MV})<4.0$ \\
Second break  & $5.0<\log (R_{\rm br,2}/{\rm MV})<7.0$ \\
\hline\hline
\multicolumn{2}{|c|}{Normalization parameters} \\
\hline\hline
Carbon  & $q_0^{\rm C}$      \\
Lithium  & $\xi_{\rm Li}$      \\
Beryllium  & $\xi_{\rm Be}$      \\
Oxygen  & $q_0^{\rm O}$      \\
\hline\hline
\end{tabular}
\end{center}
\caption{The propagation and source parameters used in our scans.
}
\label{tab:params}
\end{table}

Our mock data are a collection of \texttt{GALPROP} CR simulation
results, including two sources: one from global fitting and one from feedback data. 
Ideally, the training data cannot be biased in particular parameter space regions, 
and the best choice is to use an ideal uniform scan with perfect coverage. 
Nevertheless, performing an ideal uniform scan is not feasible, 
especially in a high-dimensional parameter space.  
Therefore, we utilize the MCMC sampling tool to explore the high-dimensional parameter space, and the collected samples are our initial mock dataset. 
Our machine can learn and recognize 
the correlation between CR model parameters and the CR spectra with a mock dataset. 
Hence, the initial MCMC scan is our first kind of mock dataset.

Because MCMC sampling may not cover the parameter space perfectly, we also implement feedback to the dataset. 
We collect those feedback data from the validation procedure results and use them as our second type of mock dataset. 
In the following, we briefly summarize these two different mock datasets.
Note that the detailed description for the generation of the feedback dataset will be given in Sec.~\ref{sec:feedback}.

The first CR mock dataset is collected from MCMC sampling tool together with the \texttt{GALPROP} simulations. 
We use the Metropolis-Hastings algorithm to generate Markov chains. 
Based on the latest CR data (from ACE, Voyager, and AMS-02), 
our global fitting was performed by using the package \texttt{CosRayMC}~\cite{Yuan:2017ozr},
which engages external \texttt{GALPROP} code (v54) for CR propagation simulation.
Considering the coverage deficiency of MCMC scans,
our training data should not strongly depend on the current dataset.
Therefore, we perform scans with a $\chi^2$ that is a factor of $10$ less than actual data, namely,
\begin{equation}
    \chi_{\rm scan}^2=\frac{\chi_{\rm tot}^2}{10}, 
    \label{eq:chisq}
\end{equation}
where $\chi_{\rm tot}^2$ sums all $\chi^2$ values from ACE, Voyager, and AMS-02 data.
Moreover, $\chi_{\rm scan}^2$ is used to drive the direction of the scan.
We expect that future experimental data will provide smaller error bars, which   
will likely update the chi-square value as $\chi_{\rm scan}^2 \to\chi_{\rm tot}^2$.
This type of mock dataset is very conservative for this type of study.
One may also extend the mock data based on actual past, present, or future data.
However, our methodology will not be affected.

In the parameter scan, we vary the 14 model parameters described in Sec.~\ref{sec:propagation}.
The prior ranges for some of these parameters are given in Table~\ref{tab:params}.
However, the normalization parameters are just overall scaling factors
that are irrelevant to the shape of the spectra and
may introduce a parameter degeneracy with propagation and source parameters.
In addition, it is difficult to distinguish between the scaling factors and the data whitening factors
in machine learning.
Therefore, we treat these normalization parameters as nuisance parameters
that are only determined when computing the $\chi^2$ of the data.
The parameters included in the machine learning are 
\begin{equation}
    {\boldsymbol \theta}=\left\{ D_0, \delta, \eta, v_A, z_h, \nu_1,\nu_2, \nu_3, \log R_{\rm br,1},\log R_{\rm br,2}\right\}.
    \label{eq:modelpar}
\end{equation}
We collected $\mathcal{O}(2\times 10^5)$ mock data points for the inputs of the machine learning.

When using the machine trained only with the first CR mock dataset, 
we find that the machine cannot recognize some CR spectra. 
We understand that this may be because our MCMC scans do not provide 
sufficient coverage of the parameter space. 
Therefore, we attempt to fix this problem by using a {\it learning-predicting-examining loop}, 
called a {\it feedback loop} in this paper. 
We discuss this feedback loop in Sec.~\ref{sec:feedback}.
Finally, we collect approximately $\mathcal{O}(50,000)$ data points generated by the feedback loop.
These feedback data are our second mock data source, and  
we combine them with the initial MCMC samples as the training datasets.

In this work, we only prepare the mock data based on the CR model given in Sec.~\ref{sec:propagation}, 
albeit it is the most popular and well-accepted model in the community.
Currently, our machines can learn the relationship between this particular CR model and the energy spectra. 
If one wants to switch to a new CR model, a new MCMC scan has to be prepared. 
In other words, we must know the form of a function to find its inverse.  
Therefore, this could be a drawback until the ultimate CR model is built. 
Fortunately, the layer structure of our trained machines can be more or less fixed 
so that we can reuse the same machines to learn other CR models.

\subsection{Pseudo-data and artificial fluctuations}
\label{sec:pseudoexp}

A Gaussian distribution for the standard deviation can describe all the observable distributions,
including the systematic and statistical uncertainties.
Although the mean value of the total measurements fluctuates,
it will eventually stabilize after accumulating enough data.
However, as long as the central value and the standard deviation of the measured Gaussian distribution are given,
one can perform random sampling with the same Gaussian distribution to generate the pseudo-data.
We can treat the standard deviation as a fluctuation of the mean value.
Therefore, the method used to generate pseudo-data is called the \textit{pseudo} experiment.
Note that different experimental data distributions can give different pseudo-data.

When we use the trained machines for prediction, 
we feed arbitrary CR energy spectra into each machine.  
The generation of arbitrary CR energy spectra is similar to performing 
a pseudo-experiment. The only difference is that we randomly add artificial fluctuations to 
a simulated spectrum instead of using the actual experimental central values and errors. 
For convenience, the collected simulation data generated by \texttt{GALPROP} 
are called ``\textit{mock data}", and the ones produced by pseudo-experiments are called ``\textit{pseudo-data}", hereafter.
Simply speaking, our pseudo-data are formed by adding artificial fluctuations to mock data.

The traditional method to estimate the likelihood of the entire parameter space is to visit all areas of the parameter space with a sampling algorithm
and then compute its statistical strength $\chi^2$ accordingly.
However, this process utilizes a very large amount of computing resources.
In contrast, our machines can determine the propagation parameters
by using the inverse function with the energy spectra as the network inputs.
If taking the CR spectra directly from pseudo-data,
we can obtain the allowed CR model parameters efficiently
without using complicated sampling algorithms.
Namely, one can treat pseudo-experimental data
as a prior distribution rather than spending much time and effort
to find the parameter space with the highest likelihood probabilities.
In Sec.~\ref{sec:preparation}, we explicitly show how to obtain the network inputs from the generated pseudo-data.

In summary, ``Mock Data'' plus artificial fluctuations constitute the ``Pseudo-data''.

\subsection{Preparing for network inputs}
\label{sec:preparation}
This subsection addresses the technical details of the machines. 
It is recommended to skip this subsection for smoother reading if this is the first time reading the paper.

Note that we input the ratios of the spectra into the network. The reasons for using ratios are as follows:
i) to reduce normalization dependence, ii) to reduce systematics, and
iii) to more easily understand the correlation between the spectra.
More details are given below.

There are still three difficulties in preparing model inputs.
The first difficulty is that we have reintroduced four factors:
$q_0^{\rm C}$, $\xi_{\rm Li}$, $\xi_{\rm Be}$, and $q_0^{\rm O}$, which are only determined by minimizing the chi-square statistics of the CR data.
If there is a lack of CR experimental data, these four parameters are difficult to determine.
Moreover, their effects can degenerate with some combinatorial changes in propagation parameters.
Therefore, our machines cannot learn the features generated by these four normalization factors.
The second difficulty is that the experimental CR spectra from ACE, Voyager, and AMS02 in some energy bins are missing.
Therefore, it is problematic to directly use experimental bins as the basis of our network inputs
when we want to apply our machine to future data.
Finally, the third difficulty is that there can be some correlations
between different energy bins as well as different CR spectra.
CNNs might need a deeper network and spend more time training to capture the features of spectra.
In the following section, we will address solutions to these three difficulties.

Our strategy to solve the first difficulty is to determine the four
normalization factors and remove them from the pseudo-data to match our network inputs.
When utilizing \texttt{GALPROP} simulated data, which does not involve experimental data,
these four normalization factors are somewhat arbitrary and irrelevant.
When performing the $\chi^2$ calculation, experimental data can subsequently determine these four factors.
If these factors are included in the learning process,
we find that the machine has performance difficulties
because the whitening procedure often washes out the rescaling effect.

%%%%%%%%%%%%%%%%%%%%%%%
\begin{figure}
\begin{center}
\includegraphics[width=0.7\textwidth]{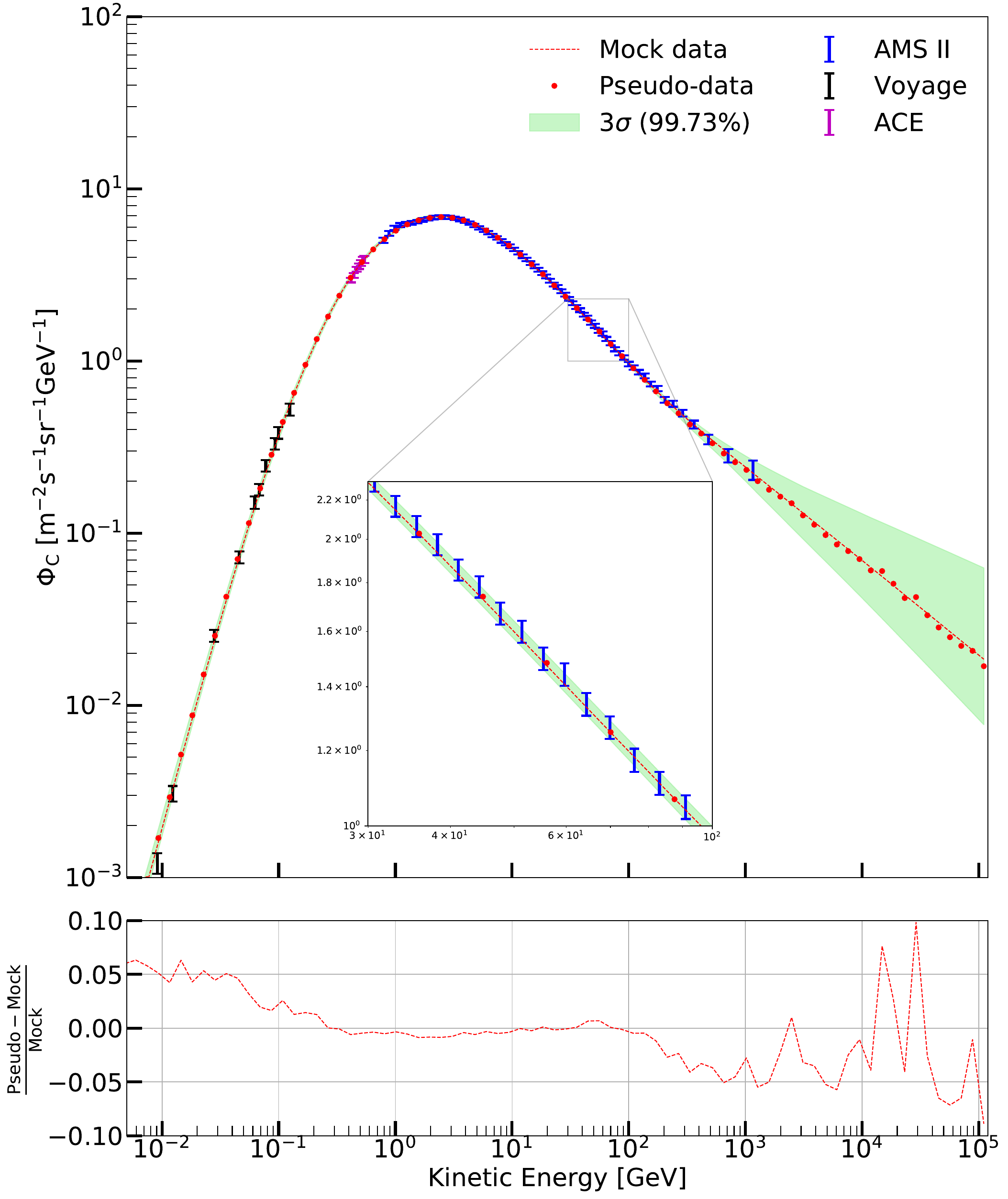}
\caption{The demodulated kinetic energy spectrum of carbon.
The solar modulation parameter $\phi$ is fixed at $696$ MV.
The experimental error bars of AMS02, Voyage, and ACE are
coloured blue, black, and purple, respectively.
The red dashed line shows a random mock spectrum.
The solid red dots represent a random pseudospectrum.
The artificial fluctuations are a maximum of 5\%.
The green band indicates the $3\sigma$ ($99.73\%$ C.L.) combined uncertainty of the latest measurements (Li, Be, B, C, and O)
from AMS02, Voyage, and ACE.
We show a zoomed-in insert for the small error bars of the AMS02 data.
The lower panel presents the relative uncertainty 
as defined by $(\text{Pseudo Data} - \text{Mock Data})/\text{Mock Data}$.
\label{Fig:fluxC}}
\end{center}
\end{figure}
%%%%%%%%%%%%%%%%%%%%%%%%%%%%%%%%%%%%%

We can fix the second difficulty (missing data in some energy bins of
CR spectra) as follows.
Instead of generating pseudo-data on the actual experimental energy bins,
we generate the pseudo-data based on new energy grids.
Specifically, we randomly pick a spectrum $\Phi^\prime(E)$ from
$4\sigma$ allowed mock data.\footnote{We expect that all future observational data
will more or less fall into the range of $\pm 4 \sigma$ allowed mock data. }
We use this selected $\Phi^\prime(E)$ as the new central value, but
the averaged ratio of the experimental error bar to the central value is the
size of the artificial fluctuation. Considering a total of $398$ experimental data points
from ACE, Voyager, and AMS02, the averaged uncertainty is approximately $5\%$ of the central value.
Hence, we adopt at most $5\%$ as our artificial fluctuation to add to the selected $\Phi^\prime(E)$ and
the new spectrum $\Phi^{\rm pseudo}$ is the actual pseudospectrum used in this study.
The new total chi-square measure becomes
\begin{equation}
\chi_{\rm tot}^2 =\sum_{i,j} 
\left( \frac{\Phi^{\rm pseudo}_{i,j} - \bar{\Phi}_{i,j}}{\sigma_{i,j}} \right)^2, 
\label{eq:pseudo_chisq}
\end{equation}
where $i$ proceeds to use experimental data from 
AMS02, Voyage, and ACE,
but $j$ is the energy bin index.
This treatment also prevents miscounting the weight for specific energy bins
when combining data from different experiments.
In Fig.~\ref{Fig:fluxC}, we show the demodulated kinetic energy spectrum of carbon.
The solar modulation parameter $\phi$ is fixed at $696$ MV.
The green band indicates the $3\sigma$ uncertainty obtained
by the latest experimental data from AMS02, Voyage, and ACE.
The $84$ bins of the pseudospectrum (red dots) are evenly located in
the kinetic energy range $10^{-3}\gev<E<1.1\times 10^{5}\gev$.
We can see that the pseudospectrum closely resembles the mock spectrum
in the small energy region where experimental error bars are small.
However, fluctuations become more significant in the region with no experimental data.
We explicitly show the relative difference between the pseudo and mock data in the bottom panel.
We refer readers to Appendix~\ref{app:CRflux} for the kinetic energy spectra of Li, Be, B, and O considered in this work.

\begin{table}[t]\footnotesize
\begin{center}
%%%%%%%%%%%%%%%%%%%%%%%%%%%%%%%%%%%%%%%%%%%%%%%%%%%%%%%%%%%%%%%%%%%%%%%%%%%%%%%%
\begin{tabular}{c|cc}
\hline\hline %\centering
%%%%%%%%%%%%%%%%%%%%%%%%%%%%%%%%%%%%%%%%%%%%%%%%%%%%%%%%%%%%%%%%%%%%%%%%%%%%%%%%
\multicolumn{3}{c}{\bf{Ratio spectra}}   \\
%%%%%%%%%%%%%%%%%%%%%%%%%%%%%%%%%%%%%%%%%%%%%%%%%%%%%%%%%%%%%%%%%%%%%%%%%%%%%%%%
\hline\hline
% %%%%%%%%%%%%%%%%%%%%%%%%%%%%%%%%%%%%%%%%%%%%%%%%%%%%%%%%%%%%%%%%%%%%%%%%%%%%%%%%
Lithium  & $\mathcal{N}_{\rm Li}$/$\mathcal{N}_{\rm C}$ & $\mathcal{N}_{\rm Li}$/$\mathcal{N}_{\rm O}$ \\
\hline
Beryllium  & $\mathcal{N}_{\rm Be}$/$\mathcal{N}_{\rm C}$ & $\mathcal{N}_{\rm Be}$/$\mathcal{N}_{\rm O}$ \\
\hline
Boron  & $\Phi_{\rm B}(E)$/$\mathcal{N}_{\rm C}$ & $\Phi_{\rm B}(E)$/$\mathcal{N}_{\rm O}$ \\
\hline
Carbon  & \multicolumn{2}{c}{ $\mathcal{N}_{\rm C}$ = $\Phi_{\rm C}(E)$/$\Phi_{\rm C}^{110}$ } \\
\hline
Oxygen  & \multicolumn{2}{c}{ $\mathcal{N}_{\rm O}$/$\Phi_{\rm C}^{110}$ } \\
\hline 
\hline
\end{tabular}
\caption{\sl The ratio of spectra to include the correlations between different CR elements. 
We use a range of kinetic energy $E$ from $10^{-3}\gev$ to $1.1\times 10^{5}\gev$ with 84 bins on a logarithmic scale.
The rows represent the normalized inputs of lithium, beryllium, boron, carbon, and oxygen given by mock data,
where $\mathcal{N}_{\rm Li}$ = $\Phi_{\rm Li}(E)/\xi_{\rm Li}$,  
$\mathcal{N}_{\rm Be}$ = $\Phi_{\rm Be}(E)/\xi_{\rm Be}$ and 
$\mathcal{N}_{\rm O}$ = $\Phi_{\rm O}(E)\times q_0^{\rm C}/q_0^{\rm O}$.  
The energy spectrum of lithium, beryllium, boron, carbon, and oxygen are presented by $\Phi(E)$.    
The normalization $\Phi_{\rm C}^{110}$ is taken from the carbon flux with energy equal to $110\gev$.}
\label{tab:Corr}
\end{center}
\end{table}
%%%%%%%%%%%%%%%%%%%%%%%%%%%%%%%%%%%%%%%%%%%%%%%%%%%%%%%%%%%%%%%%%%%%%%%%%%%%%%%%  

For the third difficulty, we can help the machines capture the features
of spectra when introducing some known correlations. 
Implementing the correlation between different energy bins or CR spectra is more complicated. 
Their correlation will depend on CR physics and cannot be directly observed from the experimental data. 
Using a spectrum from mock data as the central value of pseudo-data can automatically introduce the correlation between different energy bins. 
However, the correlation between spectra is nontrivial and depends on the propagation rather than the source parameters. 
This type of correlation can be simply identified by determining the ratios of the spectra of any two elements.
As shown in Table~\ref{tab:Corr}, we therefore introduce several ratios  
where $\mathcal{N}_{\rm Li}$ = $\Phi_{\rm Li}(E)/\xi_{\rm Li}$,  
$\mathcal{N}_{\rm Be}$ = $\Phi_{\rm Be}(E)/\xi_{\rm Be}$ and 
$\mathcal{N}_{\rm O}$ = $\Phi_{\rm O}(E)\times q_0^{\rm C}/q_0^{\rm O}$.
We note that these ratios can not only introduce correlations but can also significantly improve the performance of learning.
Hence, in this work, we use the ratios of spectra in Table~\ref{tab:Corr} as pseudo-data and inputs of our machines.

Last, we would like to comment on our usage of pseudo-data.
We design our networks for taking completed spectra as inputs,
with energies from $10^{-3}\gev$ to $1.1\times 10^{5}\gev$ with 84 bins on a logarithmic scale.
In principle, one can insert any shape of spectra, but
their energies must match these 84 energy bins.
These inputs give us flexibility for the exploration of future CR data,
which may contain a broader energy range and better energy resolution.
However, given the shortcomings of the current CR data structure
as described previously, we have to use pseudo-data, not experimental spectra, as network inputs in this work.

Note that one can freely select some spectra from mock data as the central value of the pseudo-data, 
which agrees with experimental data at a required level. 
This technique is similar to the Bayesian updating prior approach, but there are two differences; 
i) our prior distributions are a function of the energy spectra rather than CR model parameters, 
and ii) adding random fluctuations allow us to use new spectra,
which have never seen in the original mock data. 
Once future CR measurements cover all energy ranges of interest,
we can use the experimentally provided spectra as the network inputs directly.

\subsection{Machines for inverting propagation}\label{sec:machines}

We utilize two machines to learn tasks: 
\begin{itemize}
  \item {\bf Machine A}: a CNN machine to directly learn the inverse propagation from a set of mock data to a set of model parameters.
  %and vise versa.
  \item {\bf Machine B}: a CNN machine to directly learn the inverse propagation from a set of pseudo-data to a set of model parameters.
\end{itemize}
We can see that the difference between {\bf Machines A} and {\bf B}
is the input data used for training.
While {\bf Machine A} is only able to find the solutions closest to the
mock data,
we expect that {\bf Machine B} is more flexible for the general-purpose use with
inputs of {\it any} data, even with fluctuations.

\begin{figure}
\begin{center}
\includegraphics[width=0.9\textwidth]{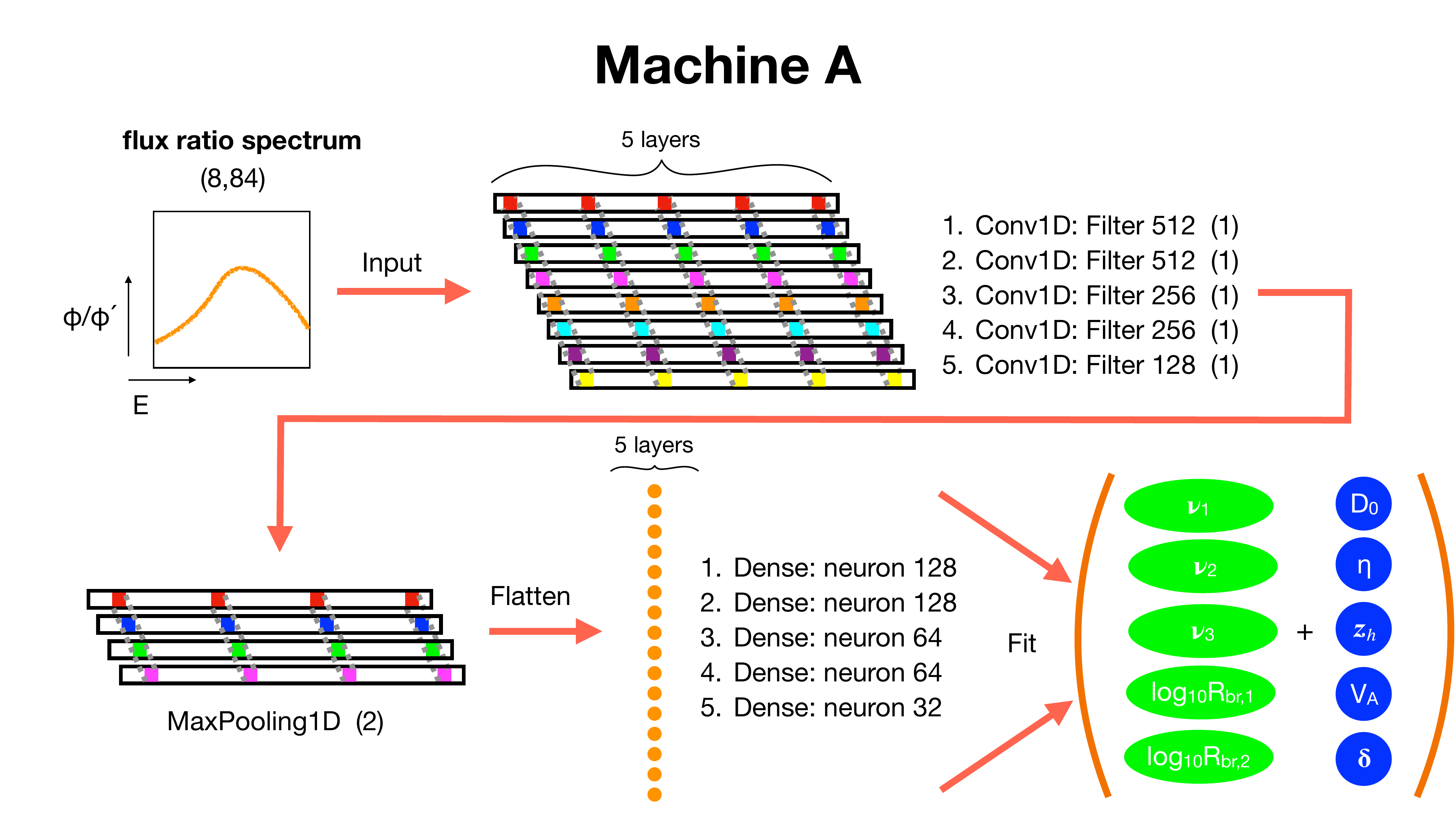}\vspace{3mm}
\includegraphics[width=0.9\textwidth]{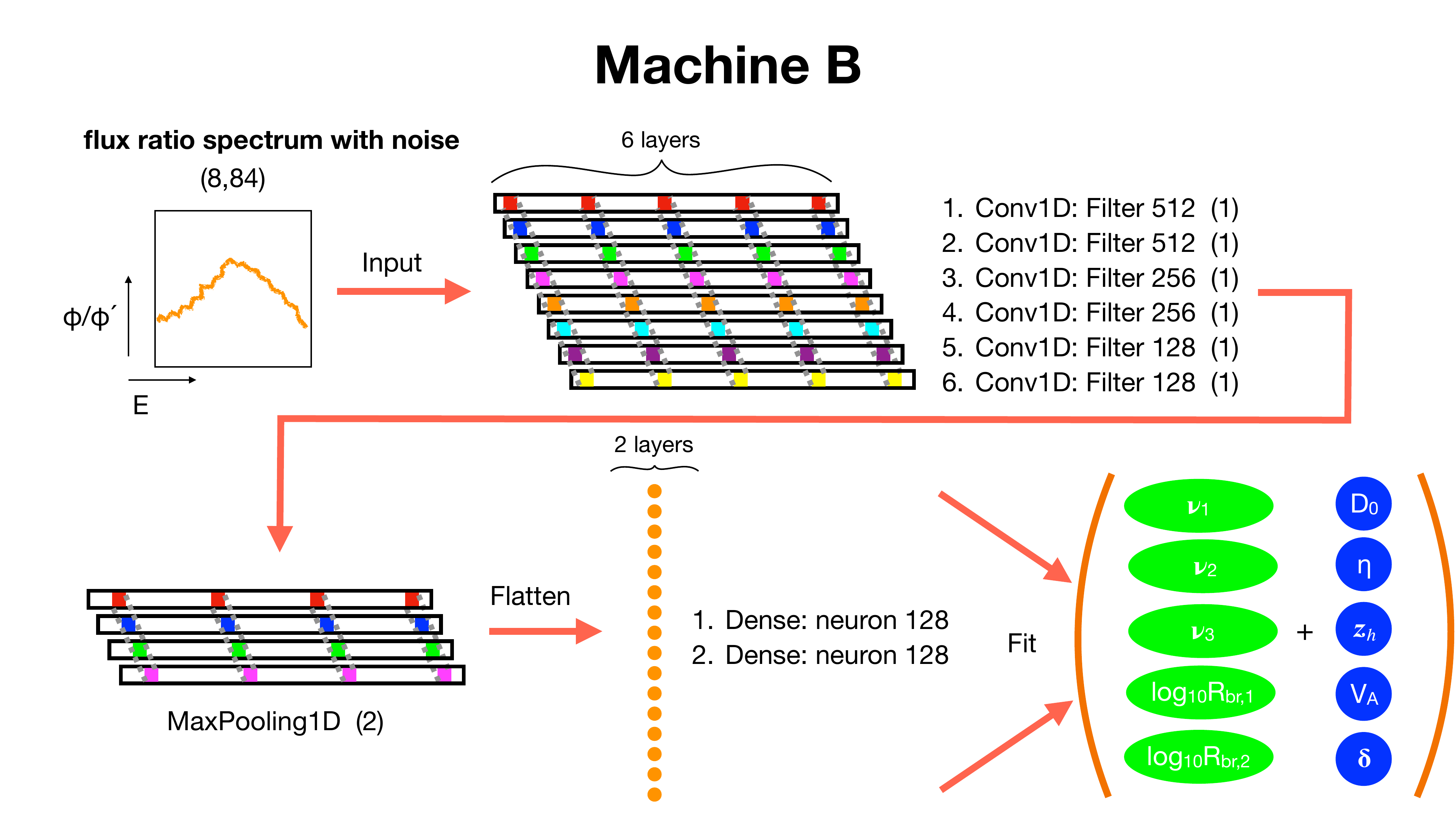}
\caption{The architectures of our {\bf Machine A} and {\bf Machine B}. 
See the text for more details.
\label{Fig:Architecture}}
\end{center}
\end{figure}

We show the architectures of {\bf Machine A} and {\bf Machine B} in 
Fig.~\ref{Fig:Architecture}. 
The input layer contains eight spectra for $84$ energy bins and the output 
layer contains $10$ propagation and source spectral shape parameters 
(the 4 normalization parameters, $q_0^{\rm C}$, $q_0^{\rm O}$, $\xi_{\rm Li}$,
and $\xi_{\rm Be}$ are not trained).  
For {\bf Machine A}, there are five one-dimensional convolutional layers,  
with $512$, $512$, $256$, $256$, and $128$ filters, accordingly.  
The argument \texttt{kernel\_size} of one-dimensional convolutional filters 
is set to $1$, the size of the maximum pooling layers is \texttt{pool\_size=2}, 
and the stride length, \texttt{strides=1}.
There are a total $5$ fully connected dense layers with $128$, $128$, $64$,
$64$, and $32$ neurons, respectively.
In the bottom panel of Fig.~\ref{Fig:Architecture}, 
we can see that the architecture of {\bf Machine B} is slightly different from that of {\bf Machine A}. 
As {\bf Machine B} is expected to have more tolerance with fluctuating input data
(this is always the case for real measurements), the learning spectra of {\bf Machine B}
include artificial fluctuations. Instead of using five one-dimensional convolutional layers,
{\bf Machine B} utilizes six one-dimensional convolutional layers, among which the additional layer contains $128$ filters.
There are only two dense layers for {\bf Machine B}, each of which contains $128$ neurons.
The exponential linear unit (ELU)~\cite{ELU} is used as the activation function in both architectures.
The last dense layer is fully connected to ten output neurons with a linear activation function in both machines.
We use Adam~\cite{Adam} as the optimizer, and the loss function is taken as $\log\cosh$.
The \texttt{Keras-2.1.6} library is used to train the machines
with \texttt{Tensorflow-1.10.1}~\cite{tensorflow2015-whitepaper} backend, on \texttt{NVIDIA Tesla P100-SXM2-16GB}.

\begin{figure}
\begin{center}
\includegraphics[width=0.49\textwidth]{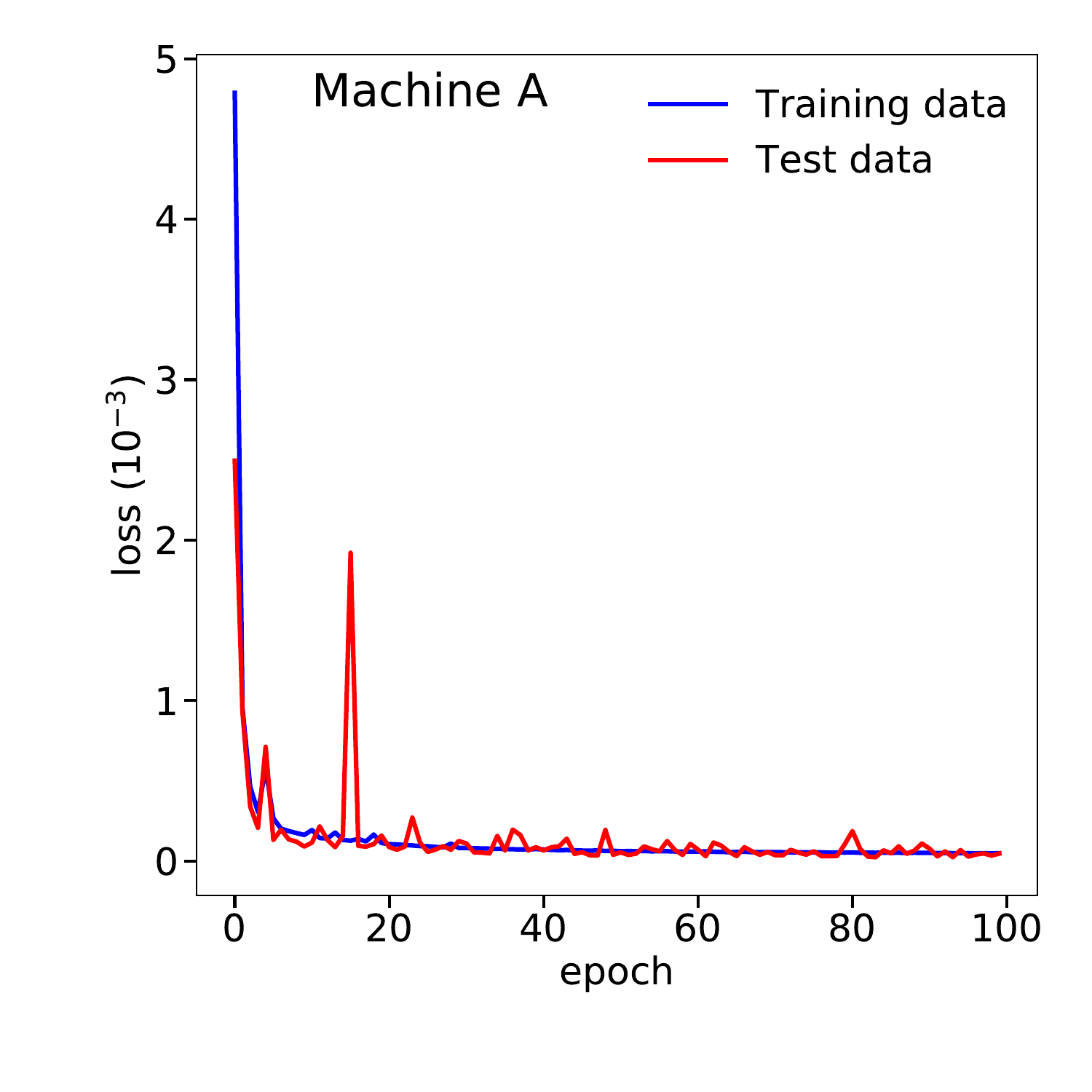}
\includegraphics[width=0.49\textwidth]{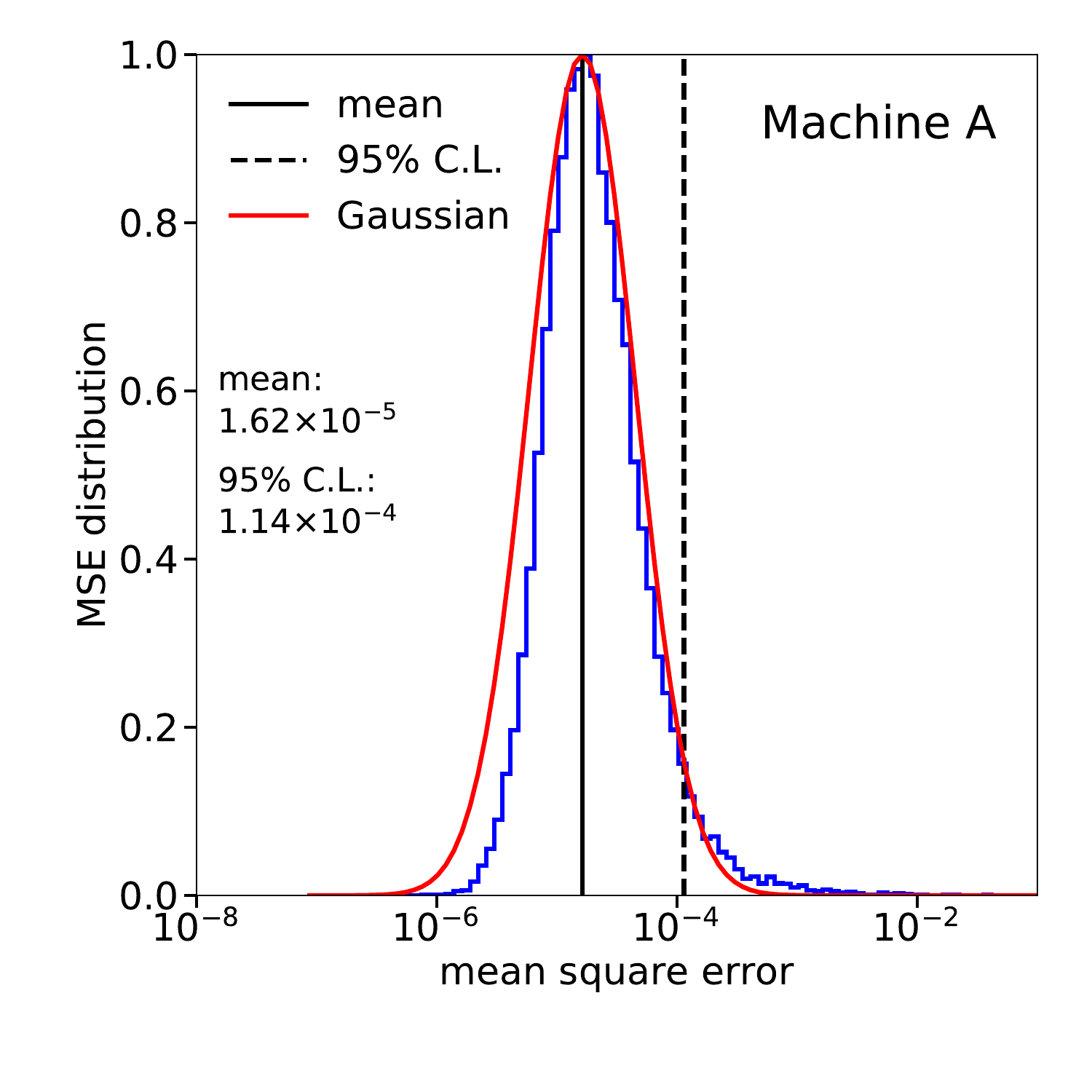}\\
\includegraphics[width=0.49\textwidth]{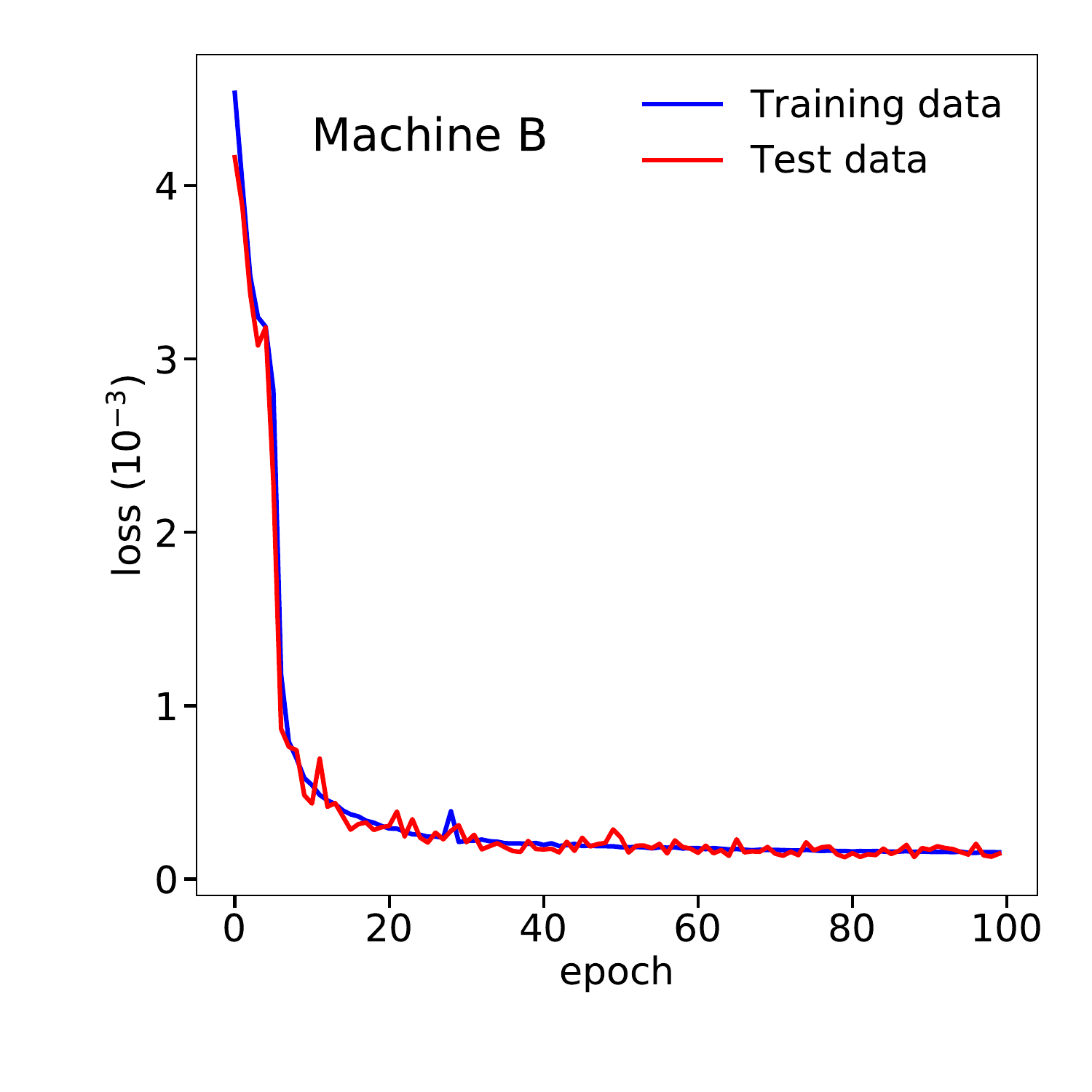}
\includegraphics[width=0.49\textwidth]{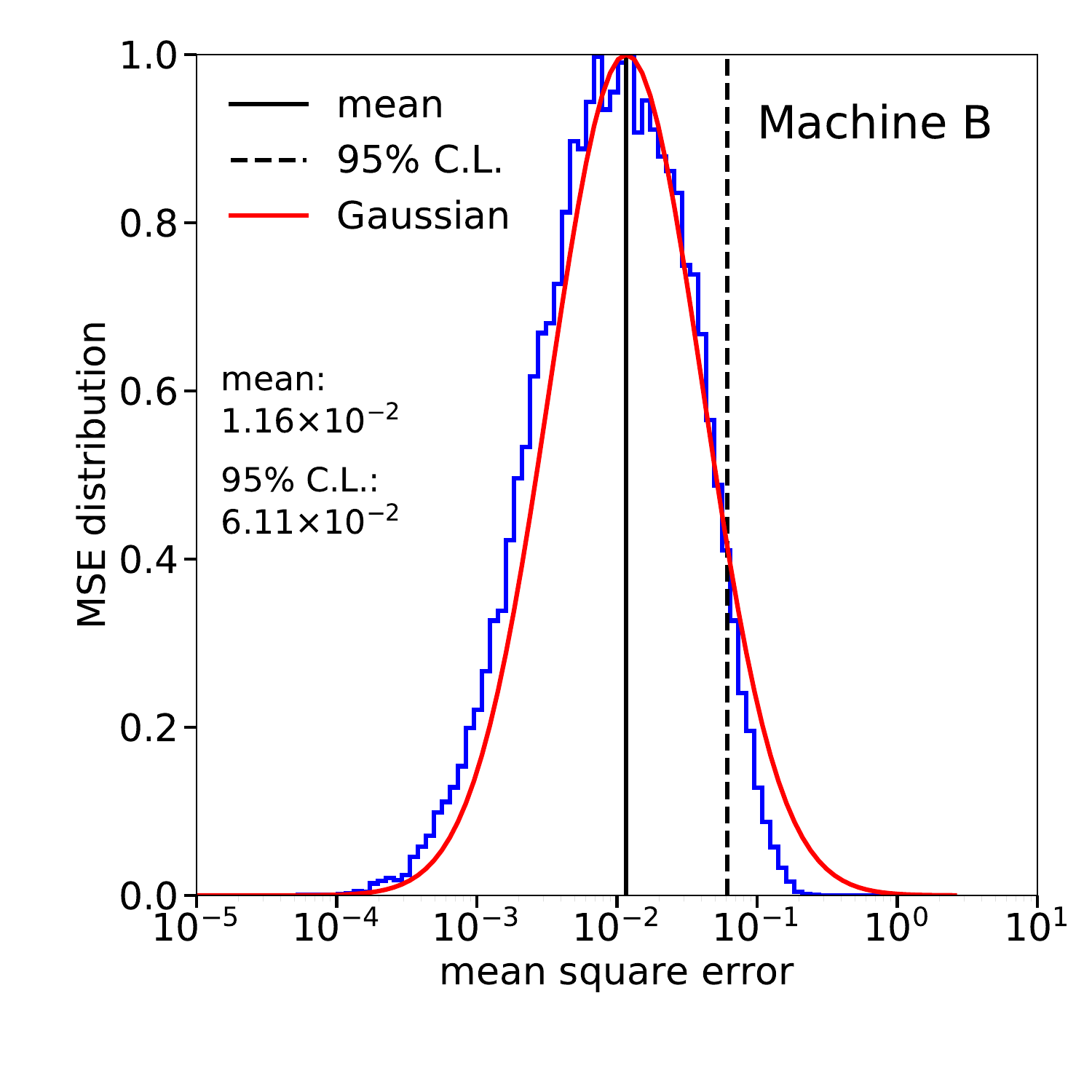}\\
\caption{
Left panels: the learning curve of training. 
Right panels: the mean square error distribution of the test data. 
The two upper figures are based on {\bf Machine A} while 
the two lower figures on {\bf Machine B}.
\label{Fig:MSE_wo_noise}}
\end{center}
\end{figure}

Both {\bf Machine A} and {\bf Machine B} can invert the cosmic-ray propagation to predict the CR model parameters 
after learning a large amount of data. As shown in Table~\ref{tab:Corr}, 
the ratios of spectra at the Earth are inputs to the machines, but the CRs model (source and propagation) parameters are outputs. 
Because of the strong correlation among the inputs, we adopt the CNN method\footnote{We have compared with 
a relatively simple Deep Neural Network, but its performance is much worse than the CNN in this case.} 
to address this type of training, such as recognizing pixels in photographs.

With 84 bins in each spectrum and a total of 8 spectra, as shown in Table~\ref{tab:Corr},
the total input degrees of freedom for {\bf Machine A} and {\bf Machine B} are $84 \times  8$.
There are ten output dimensions, five propagation parameters and five source parameters, as shown in Table~\ref{tab:params}.
We perform the whitening procedure as follows. 
Except for $R_{\rm br,1}$ and $R_{\rm br,2}$, whose range is wider than two orders of magnitude,
we also whiten the other propagation parameters to be between 0 to 1.
We take the logarithm of $R_{\rm br,1}$ and $R_{\rm br,2}$ before the whitening procedure.
After the whitening procedure, we randomly divide our total input data points so that 
$90\%$ of data is used for training and $10\%$ is used for testing.
For better performance, we choose the log-cosh loss function in both machines. 
Our trained machines can predict the source spectra and propagation parameters with some necessary tuning for the hyperparameters.
The explicit setting is given in Fig.~\ref{Fig:Architecture}.

Here, we investigate the effects of adding artificial fluctuations by comparing {\bf Machine A} and {\bf Machine B}.
Moreover, adding artificial fluctuations is a method of creating spectrum ratios different from 
those of the collected simulated data (called mock data) because our pseudo-data generation still relies on the mock data.
Once artificial fluctuations are included, 
despite some instances of unexpected spectra, 
the fluctuation effects might be degenerate with the changes in model parameters.
We build {\bf Machine B}, which learns the pseudo-data
(simulated data with artificial fluctuations) as our ``control machine". 
As a ``subject machine", {\bf Machine A} is restricted to learning the simulated data
without the implementation of artificial fluctuations.
However, we always feed the same pseudo-data for predictions.
Under two different setups, one expects that 
{\bf Machine A} is unable to recognize artificial fluctuations but makes predictions based on the most similar spectral shape.
In contrast, {\bf Machine B} has some level of capability to recognize spectra with artificial fluctuations.

In Fig.~\ref{Fig:MSE_wo_noise}, we present the performance of {\bf Machine A} (upper two panels)
and {\bf Machine B} (lower two panels).
The total loss function ($\log\cosh$) decreases to the number of training epochs in both machines.
After $\sim 80$ epochs, the loss becomes a constant for both the training and test data.
We utilize a machine based on $100$ epochs of training in our work.
The two right panels only show the conventional mean squared error (MSE) distribution for the test data.
The MSE of the test data is a well-distributed Gaussian function, which implies neither overtraining nor undertraining of the data.
In the left panel, the mean and $95\%$ C.L. lines are given by the vertical solid and dashed lines, respectively.
We especially note that the MSE distribution of {\bf Machine B} has a larger width than {\bf Machine A}.
We found that {\bf Machine B} may not be able to discriminate the changes in CR parameters from artificial fluctuations.

We want to emphasize that one can reuse the same architectures of these two machines for training different CR models        
because the purpose (finding inverse functions) is identical.
Hence, time can be saved in modifying CNN layers and hyperparameters. 
Usually, the most time-consuming procedure is not the training but the machine architecture modification. 
It may be sufficient to only retrain the last few layers in some optimistic cases.

\subsection{Feedback loop}
\label{sec:feedback}

After {\bf Machine A} and {\bf Machine B} are trained,
we can use the pseudo-data as inputs to both machines.
Moreover, the outputs $\{\theta\}$ in these two machines 
can be inserted back into the \texttt{GALPROP} code for
the validation of learning as shown by the sequence
$\mkpur{(v)}\to \mkpur{(vi)} \to \mkblue{(1)}$ in Fig.~\ref{Fig:FlowChart}.
The explicit path of the feedback loop starts with the process
from $\mkpur{(i)}$ to $\mkpur{(vi)}$ before
connecting to the label $\mkblue{(1)}$ (\texttt{GALPROP} simulator).
The first machine prediction of CR model parameters in label $\mkpur{(v)}$ is based on the pseudo-data. 
The process follows the sequence 
$\mkblue{(1)}\to \mkblue{(2)}\to \mkpur{(ii)}\to \mkpur{(iii)}\to \mkpur{(iv)}\to\mkpur{(v)}$.
Once we reach the purple label $\mkpur{(v)}$ again, we can obtain the second machine prediction and 
the feedback loop for modification ends. 
The second machine prediction of CR model parameters is based on the recomputed spectra
by \texttt{GALPROP}, which uses the first machine prediction parameters as inputs. 
We note that the randomly generated pseudo-data can be different
from the one computed by using \texttt{GALPROP}, even if we use the same CR model parameters.

In this work, for each model parameter $\theta_i$ given in
Eq.~\eqref{eq:modelpar}, 
we introduce the systematic uncertainty $\tau_i$ of network prediction as   
\begin{equation}
  \tau_i= \frac{\theta_i^\prime-\theta_i}{\theta_i}. 
\end{equation}
Comparing the result returned from \texttt{GALPROP} ($\{\theta_i^\prime\}$)
and predicted by  
randomly generated pseudo-data ($\{\theta_i\}$),
the systematic uncertainties of each parameter 
introduced by machine learning can be treated as $\{\tau_i\}$. 
Based on the set $\{\theta_i\}$, we set our stop criteria as follows:   
when $95\%$ of the total predictions agree with the condition $\frac{1}{10}\sum \left|\tau_i\right| <0.1$, 
we terminate the repeated feedback loop. 
Here, we take the averaged error of the $10$ model parameters. 
Although this criterion is not extremely stringent, 
the first machine predicted model parameters by using pseudo-data as inputs    
still give very similar values to the parameters predicted by using the spectra generated by \texttt{GALPROP}.

In label $\mkpur{(vi)}$ of Fig.~\ref{Fig:FlowChart}, if we do not reach the stop criteria, we repeat the training
with updated mock data, including those generated during the modification process.
This approach of checking for modification follows
the algorithm of ``\textit{Generative Adversarial Network}''~\cite{Goodfellow:2014upx}, 
see also its application to LHC physics~\cite{Butter:2019cae}. 
Namely, instead of hiring discriminative networks to evaluate our machines, 
we verify the result by inserting predicted parameters back to the \texttt{GALPROP} simulation.

\subsection{Data processing}
\label{sec:processing}

In this subsection, we summarize the data processing and all our steps in Fig.~\ref{Fig:FlowChart}.
This blocks show all the steps adopted in our work, and the arrows represent the direction of data processing.
For example, the sequence $\mkblue{(1)}\to \mkblue{(2)}\to \mkblue{(3)}$
indicates the following operations.
First, we use \texttt{GALPROP} $\mkblue{(1)}$ to generate a set of data
including CR model parameters and spectra.
Then, we append these data to other pre-existing datasets (if any)
to form a large collection called mock data $\mkblue{(2)}$.
Finally, we send a very large amount of mock data to {\bf Machine A} for
training $\mkblue{(3)}$.
Likewise, the sequence $\mkpur{(i)}\to \mkpur{(ii)}\to \mkpur{(iii)} \to
\mkpur{(iv)} \to \mkpur{(v)}$ can be understood as follows.
From any CR experimental data, we can extract the central values and errors in $\mkpur{(i)}$.
With the updated central values and errors, we can recompute the new $\chi^2$ for our massive mock data collection $\mkblue{(2)}$.
Requiring specific statistical criteria, for example, $\delta \chi^2<33.2$ for spectra less than
$3\sigma$ CI in our study,
one can randomly select a set of model spectra from this refined $3\sigma$ bracket.
In label $\mkpur{(ii)}$, we perform a pseudo-experiment by simply
adding random artificial fluctuations that do not exceed $5\%$ to
this selected model spectra.
The newly formed spectra are pseudospectra.
We then use these pseudospectra to perform data preprocessing for
calculations of ratio spectra $\mkpur{(iii)}$
and whitening transformation $\mkpur{(iv)}$.
Eventually, our two trained machines can read their inputs given by $\mkpur{(iv)}$
and predict their corresponding CR model parameters as label $\mkpur{(v)}$.
Note that we exactly compute $\chi^2$ as Eq.~\eqref{eq:pseudo_chisq}
by using pseudospectra in $\mkpur{(ii)}$,
not the simulated model spectra in $\mkblue{(2)}$.

In this paper, we begin our work from \texttt{GALPROP} simulations,
along the sequence $\mkblue{(1)}\to \mkblue{(2)}\to \mkblue{(3)}$ for {\bf Machine A} and
$\mkblue{(1)}\to \mkblue{(2)}\to \mkblue{(3^\prime)} \to \mkblue{(4^\prime)}$ for {\bf Machine B}.
After the machines are built, one can use the machines to predict the CR model parameters
by inserting the pseudo-data following 
the sequence $\mkpur{(i)}\to \mkpur{(ii)}\to \mkpur{(iii)} \to \mkpur{(iv)} \to \mkpur{(v)}$.

We adjust our two machines by connecting the above training and prediction processes in a feedback loop.
The first machine prediction of the CR model parameters in the feedback loop is
via $\mkpur{(i)} \to\mkpur{(ii)} \to\mkpur{(iii)} \to\mkpur{(iv)} \to\mkpur{(v)}$, where the inputs are the pseudospectra. 
However, we can insert the first machine predicted model parameters
back to the \texttt{GALPROP} simulation $\mkblue{(1)}$ to recompute the actual spectra.
Then, these spectra with $<5\%$ random artificial fluctuations can be fed back to both machines
to make the second prediction of CR parameters at the label $\mkpur{(v)}$.
In this step, we can compare the first and second machine-predicted CR parameters to validate our training.
If their averaged error does not satisfy $\frac{1}{10}\sum \left|\tau_i\right| <0.1$,
we have to update the mock data collection and modify our networks.
The new data appended to the old mock data 
are the CR model parameters obtained from the first machine prediction
and its \texttt{GALPROP} simulation. 
Therefore, our verification in the feedback loop is achieved by the sequence
$\mkpur{(v)}\to\mkpur{(vi)} \to \mkblue{(1)}\to \mkblue{(2)}\to \mkpur{(ii)}\to
\mkpur{(iii)}\to \mkpur{(iv)}\to\mkpur{(v)}$.

Finally, we emphasize that the experimental data can be updated. 
As long as the physical model of the propagation and source are not altered, our trained machines are reusable. 
Overall, our machines can quickly predict the allowed model parameters by plugging in new experimental data,  
even if the data have not been seen before or during the training.

%%%%%%%%%%%%%%%%%%%%%%%%%%%%%%%%%%%%%%%%%%%%%%%%%%%%%%%%%%%%%%%
\section{Results}\label{sec:result}
%%%%%%%%%%%%%%%%%%%%%%%%%%%%%%%%%%%%%%%%%%%%%%%%%%%%%%%%%%%%%%%

In this section, we show our results with the updated chi-square metric, 
$\chi^2_{\rm{tot}}$ in Eq.\eqref{eq:chisq}. Note that our mock data are mainly 
prepared by performing the MCMC scan with $\chi^2_{\rm{tot}}/10$, and hence
the $1\sigma$ and $2\sigma$ regions based on the current data (determined by
$\chi^2_{\rm{tot}}$) could be less precise. Therefore, our only concern is the
$3\sigma$, $4\sigma$, and $6\sigma$ regions. For the parameter constraints with
$1\sigma$ and $2\sigma$ confidence levels, one can refer to Ref.~\cite{Yuan:2018lmc}.

The minimum $\chi^2_{\rm{tot}}$ is $347.0$ based on the total collected mock data.
Considering the total $14$ degrees of freedom as seen in Table~\ref{tab:params},
the confidence interval (CI) for $3\sigma$ ($99.73\%$), $4\sigma$ ($99.99\%$),
and $6\sigma$ ($\approx 100\%$) are $\delta \chi_{\rm{tot}}^2$ equal to $33.20$, $43.82$ and $69.91$, respectively.
Our presentation of parameter space for mock data is based on 
the Frequentist approach, i.e., the ``profile likelihood" method (minimum chi-square), to eliminate uninteresting parameters.
When plotting the contours in this work, we use the data generated from the 
MCMC scan and the feedback loop. However, the feedback loop data focus on the 
regions where the machines do not learn the inverse function well. Hence, utilizing 
the Bayesian marginal posterior approach is inappropriate because the prior 
distributions are unclear.

\subsection{The systematic uncertainties introduced by learning}

\begin{figure}
\begin{center}
\includegraphics[width=0.45\textwidth]{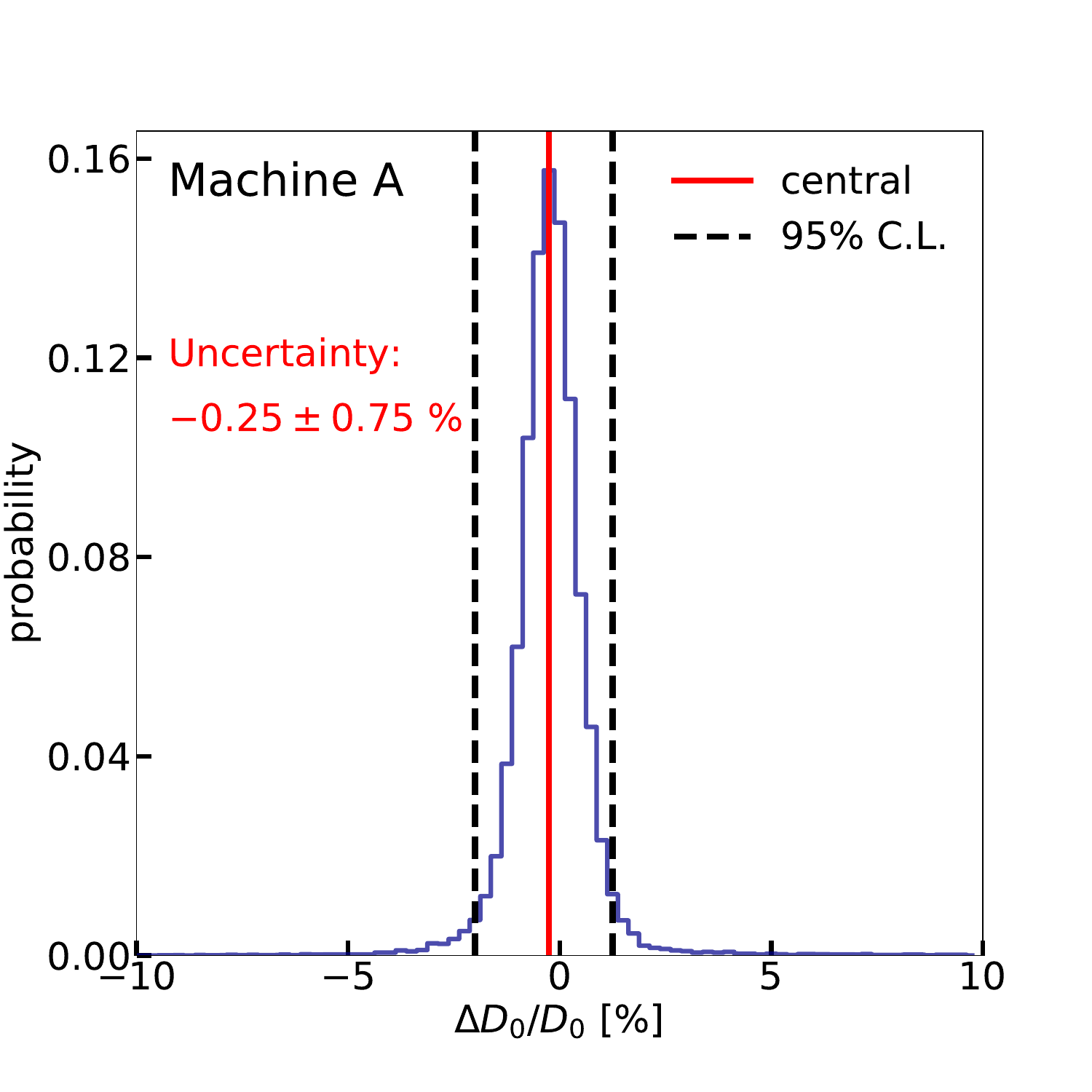}
\includegraphics[width=0.45\textwidth]{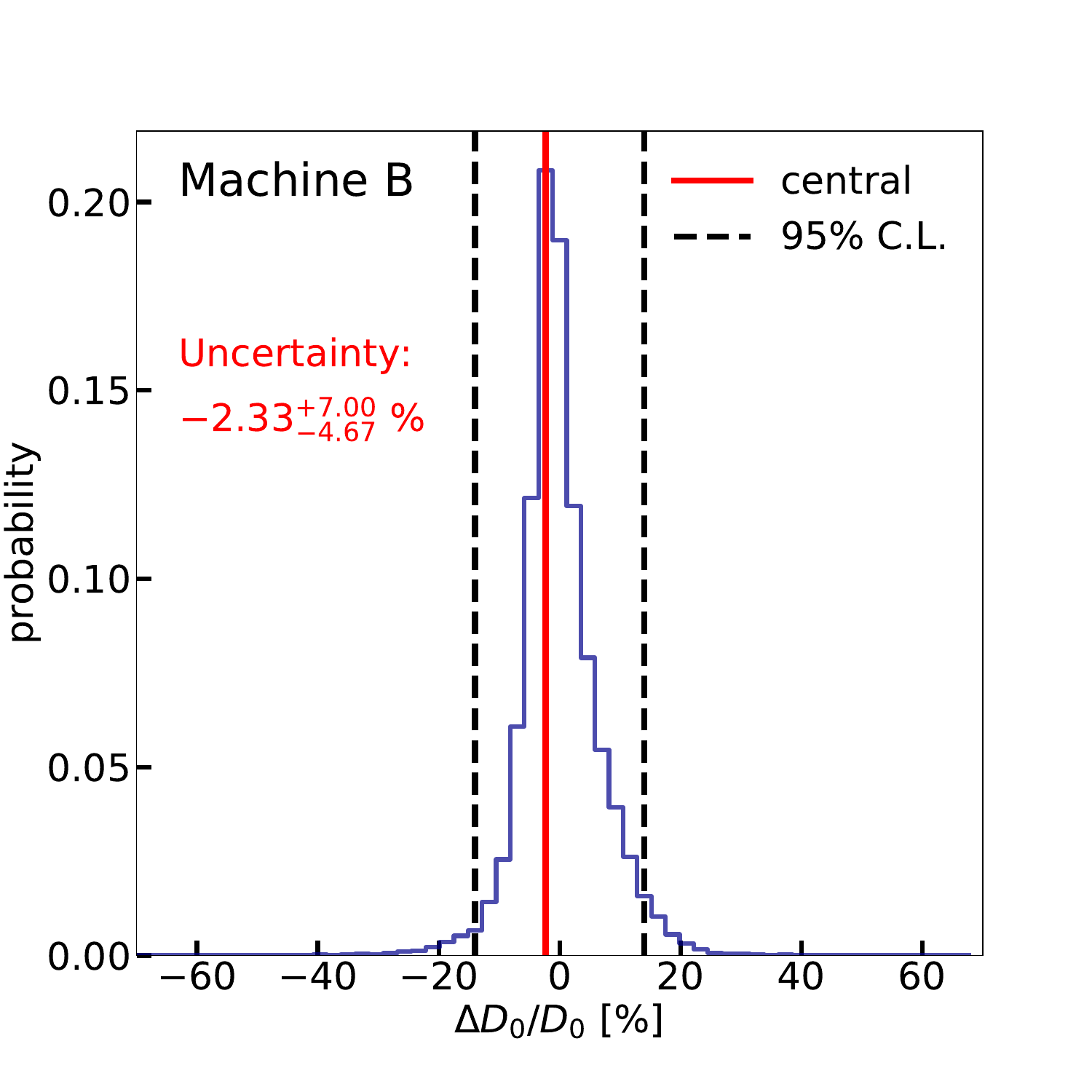}
\caption{The systematic uncertainties introduced by the learning process.
The left panel is based on {\bf Machine A}, while the right panel is based on {\bf Machine B}.
The histograms are made by using 30,000 test samples.
\label{Fig:uncertainty}}
\end{center}
\end{figure}

We first discuss the systematic uncertainties introduced by 
{\bf Machines A} and {\bf B}.
These results only depend on the mock data generated by 
\texttt{CosRayMC} with $\chi_{\rm{scan}}^2$ as given in Eq.~(\ref{eq:chisq}). 
Using this setup, we will know nothing about the current or future data, but the training is not affected.  
In Sec.~\ref{sec:cf_ml_mcmc}, we will open the box so that 
the experimental chi-square value is updated to $\chi_{\rm{tot}}^2$.

\begin{table}[htb!]
\begin{center}
\begin{tabular}{|c|c|c|}
\hline
\hline
\multicolumn{3}{|c|}{Systematic uncertainties in percentage (\%)} \\
\hline\hline
Source parameters &  Machine A & Machine B \\
\hline
$\nu_1$ & $0.00^{+3.75}_{-2.50}$ & $0.00\pm{5.00}$ \\
$\nu_2$ & $0.05^{+0.09}_{-0.14}$ & $-0.05\pm{0.15}$  \\
$\nu_3$ & $0.09^{+0.14}_{-0.09}$ & $-0.20\pm{0.20}$ \\
$\log R_{\rm br,1}$ & $0.04\pm{0.14}$ & $0.00\pm{0.30}$ \\
$\log R_{\rm br,2}$ & $-0.09\pm{0.14}$  & $0.15^{+0.30}_{-0.45}$ \\
\hline\hline
Propagation parameters &  Machine A & Machine B \\
\hline
$D_{0}$ & $-0.25\pm{0.75}$ & $-2.33^{+7.00}_{-4.67}$ \\
$\delta$ & $0.04^{+0.27}_{-0.18}$ & $-0.33\pm{0.50}$ \\
$z_h$ & $-0.50^{+1.25}_{-1.50}$ & $3.28\pm{13.11}$  \\
$v_A$ & $0.35\pm{1.05}$  & $0.00\pm{3.33}$  \\
$\eta$ & $0.00^{+0.70}_{-1.40}$ & $1.67^{+5.00}_{-1.67}$ \\
\hline\hline
\end{tabular}
\end{center}
\caption{The systematic uncertainties of different parameters introduced by learning. 
The error bars are estimated by using test data. 
}
\label{tab:uncertainties}
\end{table}

In Fig.~\ref{Fig:uncertainty}, histograms of the $\Delta D_0/D_0$
systematic uncertainties introduced in the learning process of {\bf Machine A} (left panel) and {\bf Machine B} (right panel) 
are created from the test data. 
Both histograms are well-distributed Gaussian with central values close to zero, 
and the $1\sigma$ ($68\%$ CI) regions are given by
$-0.0025\pm 0.0075$ for {\bf Machine A} but $-0.0233^{+0.0700}_{-0.0467}$ for {\bf Machine B}. 
Clearly, {\bf Machine B} has more significant uncertainties due to the added artificial fluctuations. 
Therefore, adding $5\%$ artificial fluctuations can contaminate 
the propagation inverter.

We summarize the systematic uncertainties of learning for the rest of the parameters in Table~\ref{tab:uncertainties}. 
Most of the parameters can be predicted very well, with the central values close to zero and small error bars.  
However, the networks do not predict the diffusion coefficient $D_0$ and the height of diffusion zone $z_h$ precisely, 
due to the weaker correlation between $D_0$ and $z_h$ at large values, as shown in Fig.~\ref{Fig:phasespace}. 
Typically the secondary-to-primary ratios of CR nuclei determine the ratio of $D_0/z_h$~\cite{Maurin:2001sj}. 
Indeed, our machines learn this effect. 
This scaling may be broken when $z_h$ is so large that it is comparable to the boundary of the propagation halo. 
Therefore for $z_h>10\kpc$, the performance of our machines is not as good as that in other regions.

\subsection{Comparison between the network predictions and the reweighted MCMC method}

\label{sec:cf_ml_mcmc}
\begin{figure}[htb!]
\begin{center}
\includegraphics[width=0.45\textwidth]{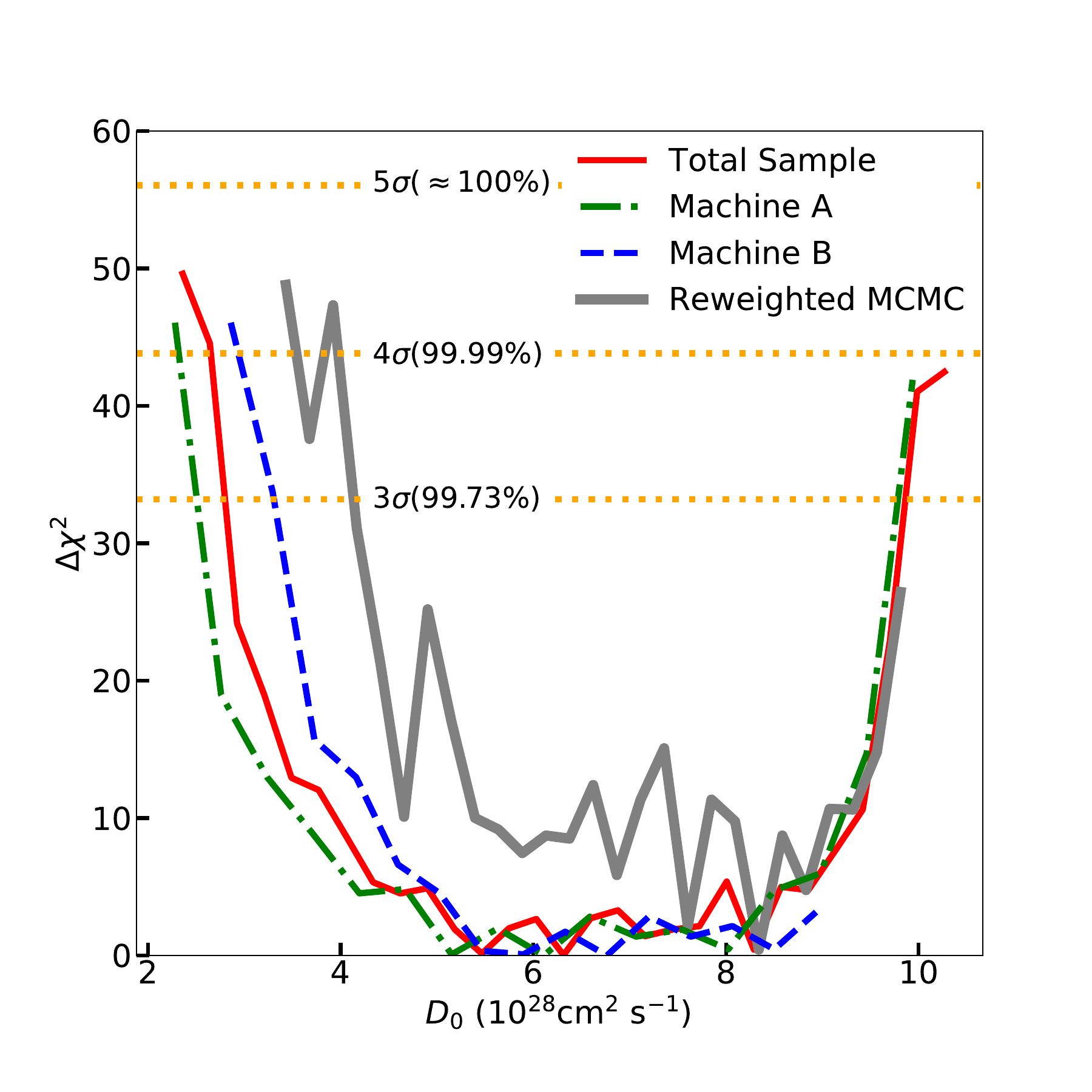}
\includegraphics[width=0.45\textwidth]{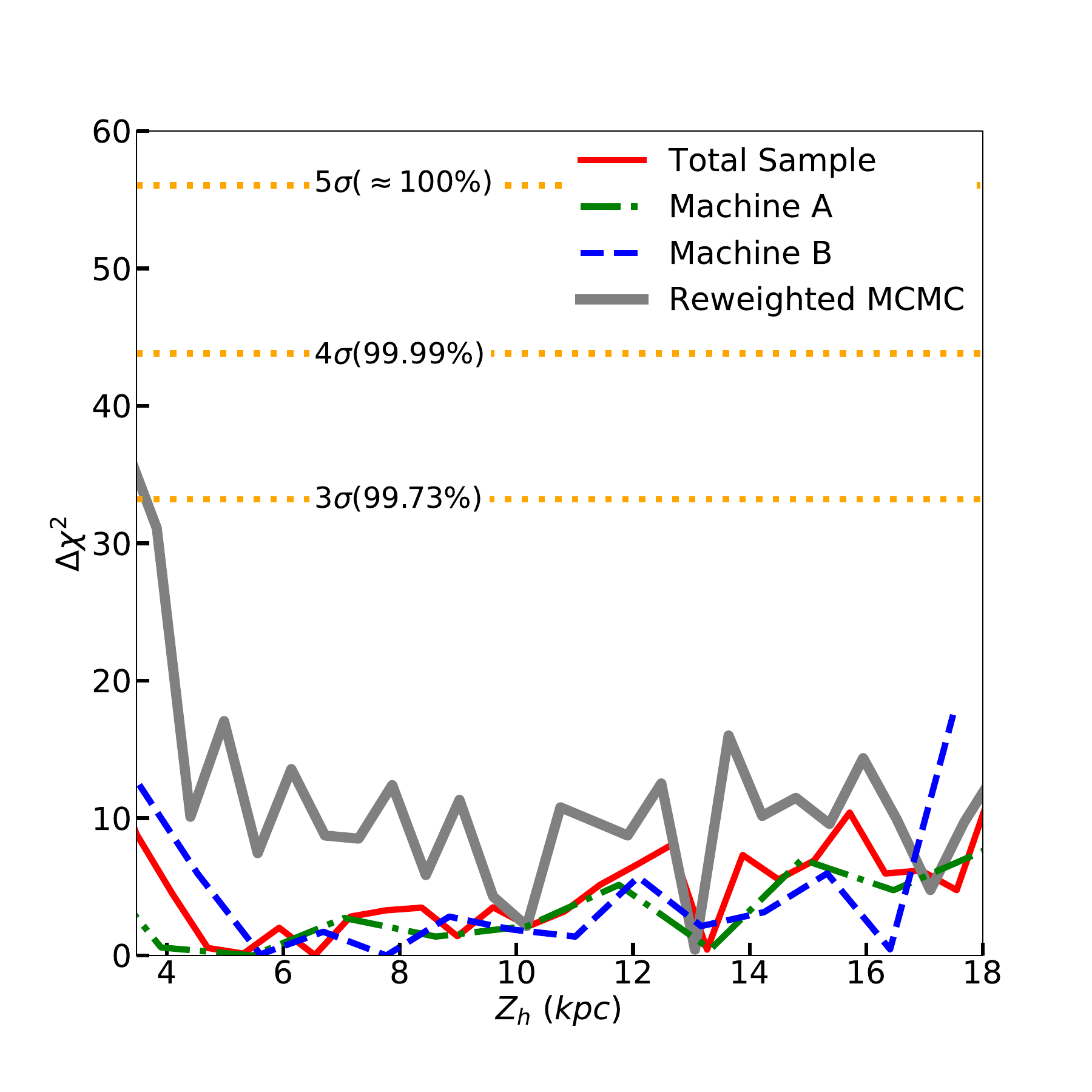}\\
\includegraphics[width=0.45\textwidth]{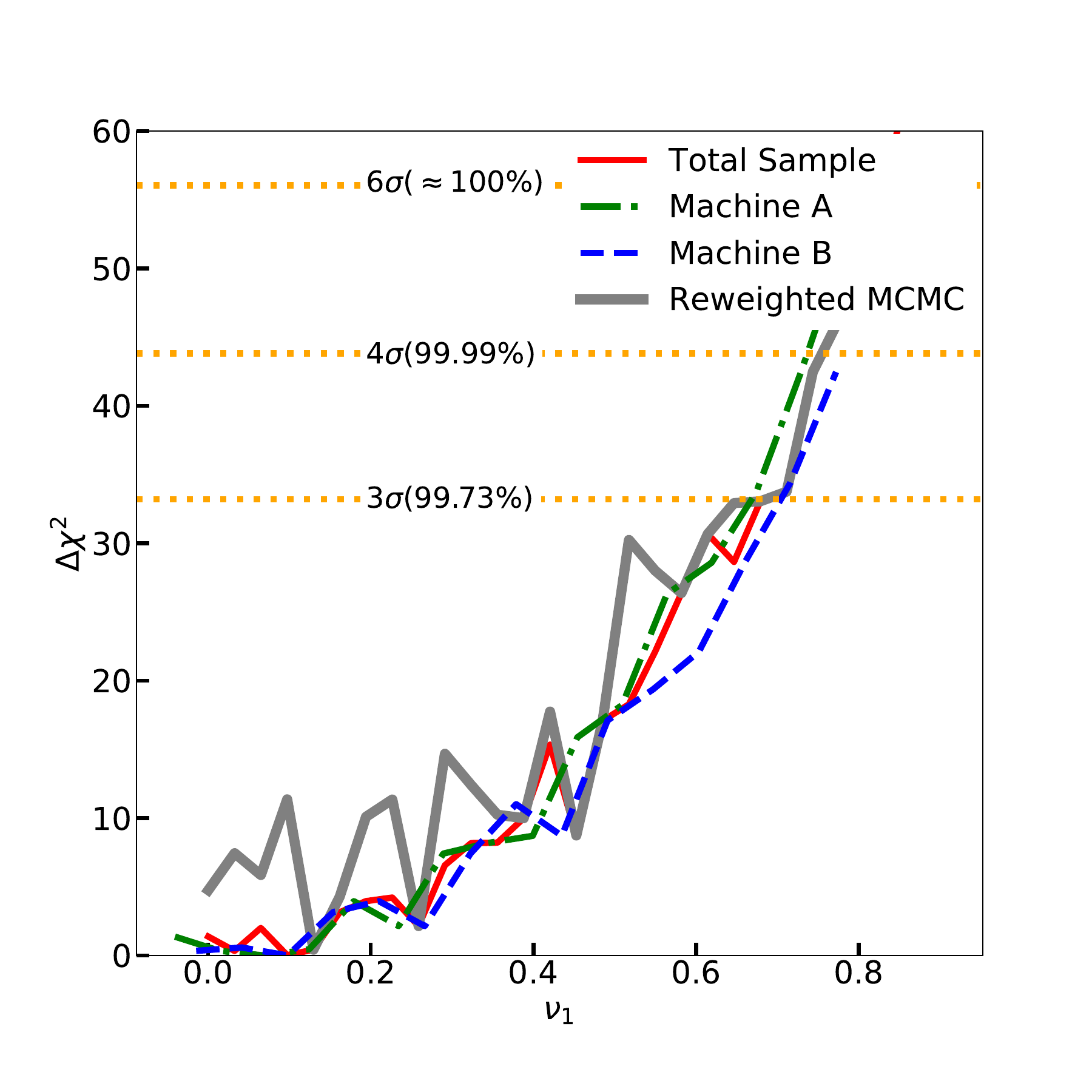}
\includegraphics[width=0.45\textwidth]{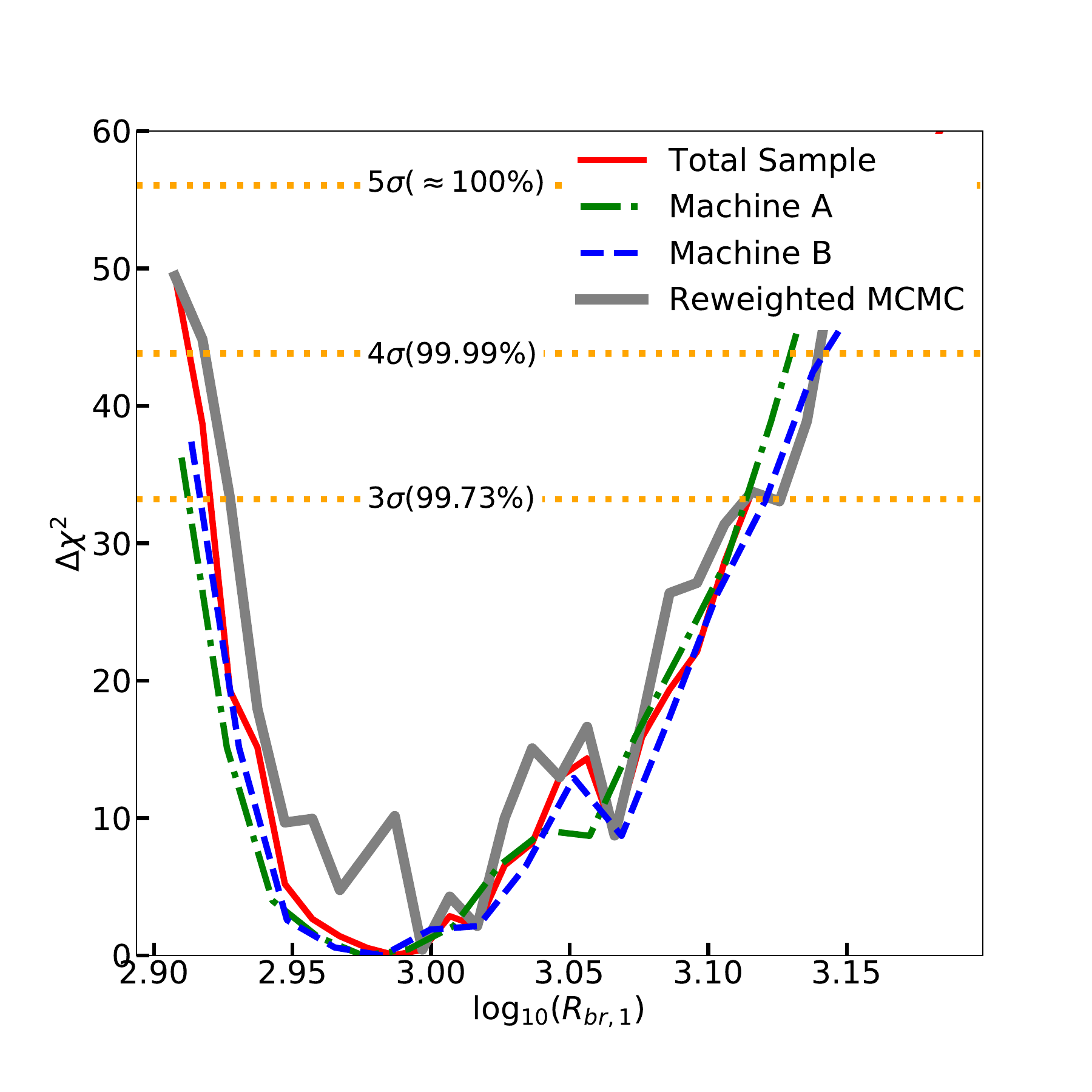}
 \caption{
 \label{fig:1Dlike}
The one-dimensional $\Delta\chi^2$ distribution for $D_0$ (upper left), $Z_h$ (upper right),
$\nu_1$ (lower left), and $R_{br,1}$ (lower right).
The $\Delta\chi^2$ distributions are projected to our CR parameter of interest,
and the rest are profiled using the profile likelihood method.
The red solid, grey thick solid, blue dashed, and green dash-dotted curves represent
the $\Delta\chi^2$ distribution of the total training samples, 
the initial MCMC scan, the predictions of {\bf Machine A} and the predictions of {\bf Machine B}, respectively.}
 \end{center}
 \end{figure}

In Fig.~\ref{fig:1Dlike}, we present the one-dimensional $\Delta\chi^2$ distribution 
for $D_0$ (upper left), $Z_h$ (upper right), $\nu_1$ (lower left), and $R_{br,1}$ (lower right).
We use the profile likelihood method to remove other unwanted parameters.
The solid red lines are based on the training samples, including the collected data from the initial MCMC and the feedback loops.
The grey thick solid lines represent the $\Delta\chi^2$ distribution based on the initial MCMC scan.
We collect the initial MCMC data using the likelihood function given in Eq.~\eqref{eq:chisq}.
For comparison, we randomly take $5\times 10^4$ pseudospectra for the network predictions of 
{\bf Machine A} (blue dashed lines) and {\bf Machine B} (green dash-dotted lines).
The orange horizontal dotted lines show the CI for $3\sigma$, $4\sigma$, and $5\sigma$.

Generally, the network predictions perform well but
{\bf Machine A} perfectly captures the features of the $\Delta\chi^2$ distribution,
even though the $\Delta\chi^2$ distributions are previously unseen information.
Importantly, we can see that both {\bf Machine A} and {\bf Machine B} manage to find the trends from the training data,
and they yield smoother curves than the MCMC method and our training samples.
As expected, the feedback loop self-corrects the problems introduced by the poor coverage of the parameter space.
In the initial MCMC scans, the networks cannot precisely predict the parameter space using extrapolation.
Because of the condition $\frac{1}{10}\sum \left|\tau_i\right| <0.1$,
we send extrapolating predictions back to the feedback loop
even if they do not exist in the initial MCMC scans.
Hence, the total training samples (red lines) are more smoothly distributed.
Note that if our networks would not precisely predict the future data,
they may extrapolate instead of interpolating the training samples and
the feedback loop can be applied again for self-correction.
For the sake of completeness, we also give the $\Delta\chi^2$ distributions of 
all other CR model parameters in Appendix \ref{sec:1d2dlike}.

%%%%%%%%%%%%%%%%%%%%%%%
\begin{figure}[htb!]
\begin{center}
\includegraphics[width=0.45\textwidth]{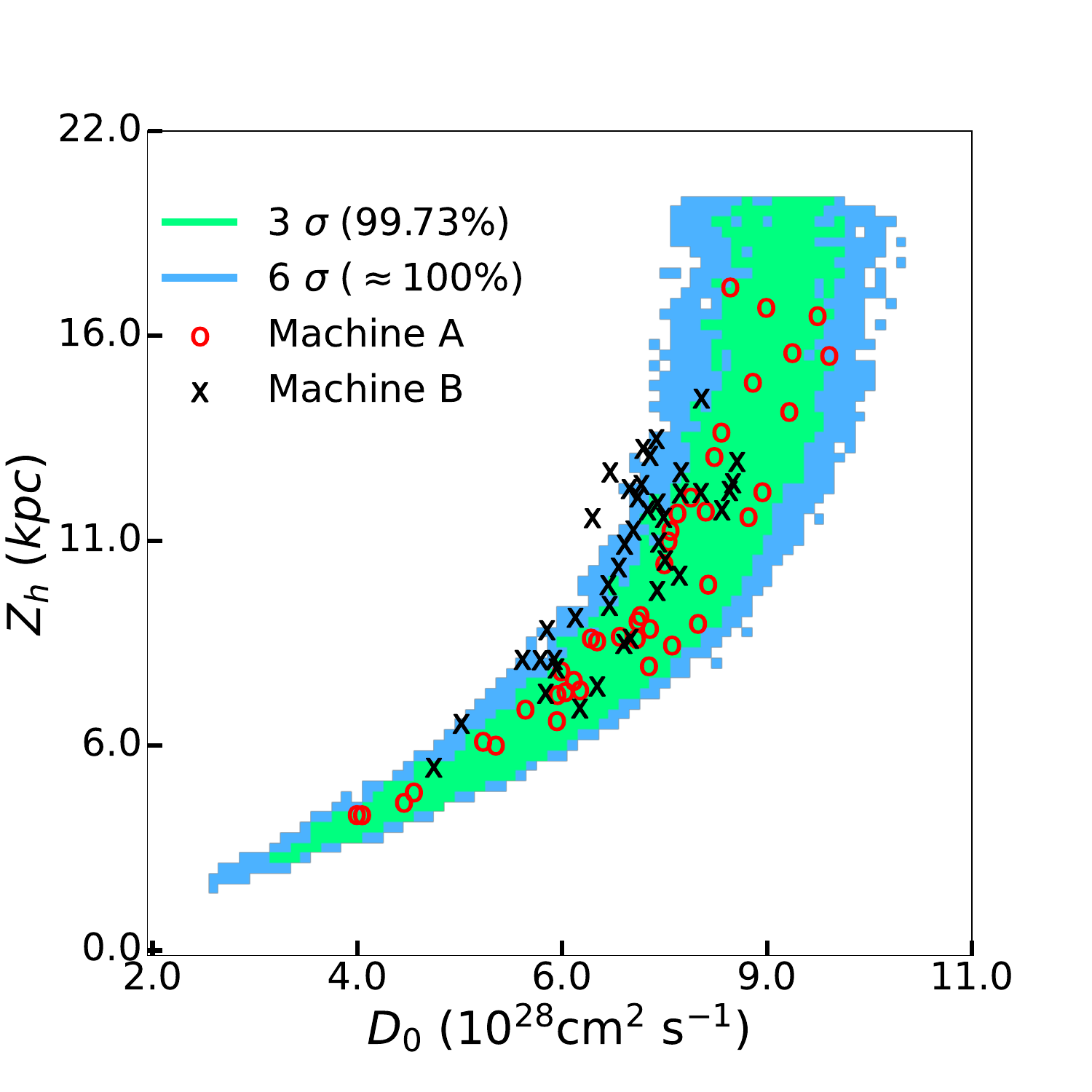}
\includegraphics[width=0.45\textwidth]{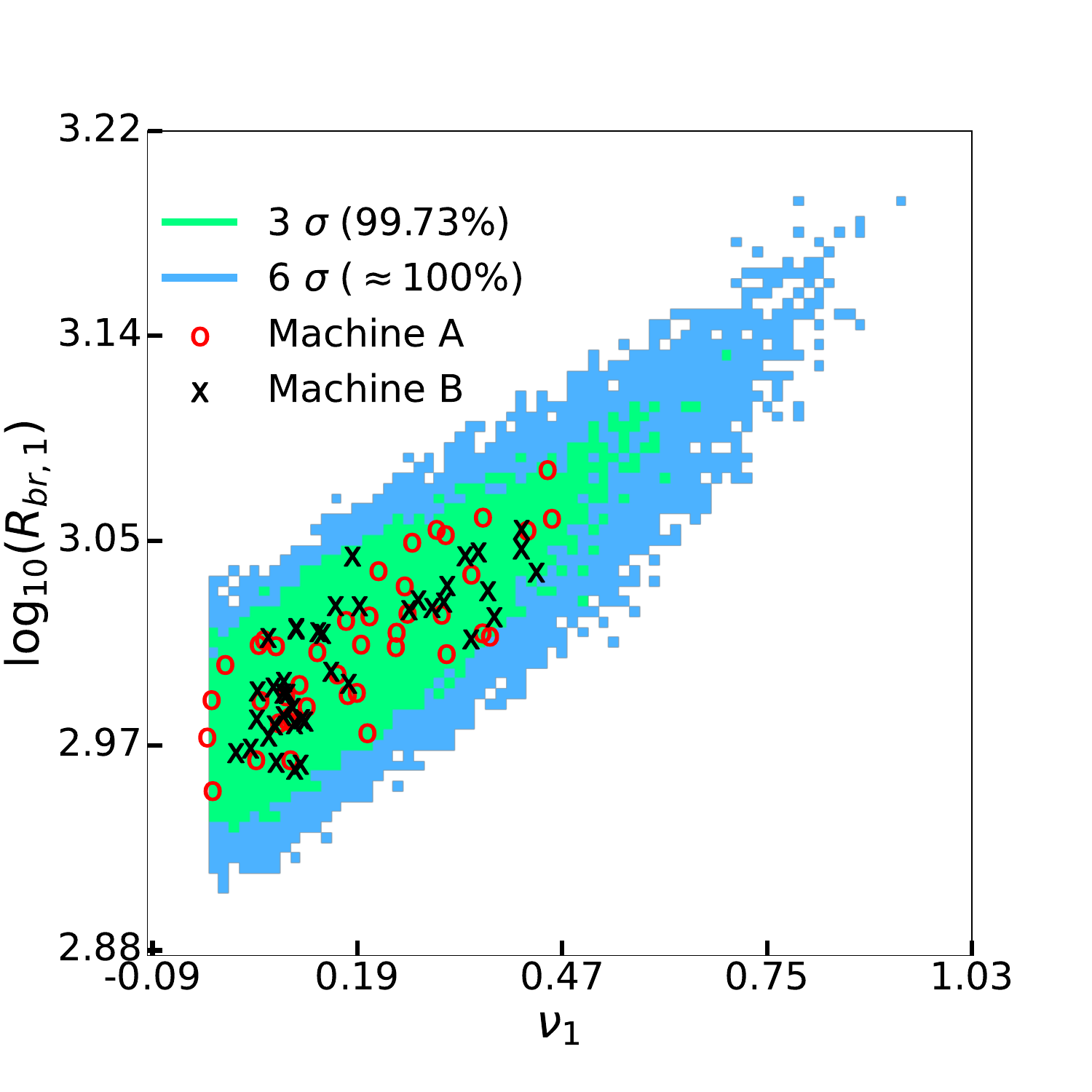}
\caption{Two-dimensional allowed parameter space.
The left panel is the allowed region projected on the ($Z_h$, $D_0$) plane, while  
the right panel is for the source parameters $R_{br,1}$ and $\nu_1$.
Based on the latest chi-square metric $\chi^2_{\rm tot}$,
the green region is the $3\sigma$ confidence interval 
and the blue region is the $6\sigma$ confidence interval.
The red circles and black crosses are 100 network predictions 
by {\bf Machine A} and {\bf Machine B}, respectively.
\label{Fig:phasespace}}
\end{center}
\end{figure}

%%%%%%%%%%%%%%%%%%%%%%%%%%%%%%%%%%%%%

In Fig. \ref{Fig:phasespace}, the propagation parameters $D_0$ and $z_h$ (left panel)
and source parameters $R_{br,1}$ and $\nu_1$ (right panel) are projected on
the two-dimensional contour plots.
We also illustrate the predicted scatter points by both machines 
with the pseudospectra within required a $4\sigma$ region 
($3\sigma$ mock data plus $1\sigma$ artificial fluctuations).
The inner contour is the $3\sigma$ region ($\delta \chi_{\rm{tot}}^2=33.20$)
while
the outer contour is the $6\sigma$ region ($\delta \chi_{\rm{tot}}^2=69.91$).
We take $100$ representative samples of {\bf Machine A} (red circles) and {\bf Machine B} (black crosses).
There is no dependence on $D_0$ at large values of $D_0$ and $z_h$. 
Therefore, both {\bf Machine A} and {\bf Machine B}, but especially for {\bf Machine B}, cannot perform well. 
However, because the correlation between $R_{br,1}$ and $\nu_1$ is clear,
both machines can predict the contours quite well. 
For other combinations of CR model parameters, refer to Appendix~\ref{sec:1d2dlike}.

\begin{table}[h!]
\begin{center}
\begin{tabular}{|c|c|c|c|c|}
\hline
\hline
\multicolumn{5}{|c|}{Prediction capabilities after updating to $\chi^2_{\rm{tot}}$} \\
\hline\hline
 &  MCMC (re-weighted) & {\bf Machine A} & {\bf Machine B} & Check Machine\\
\hline
% \# of data & 309445 & 40723 & 8382 & 7073  \\
 CI $\leq$ 4 $\sigma$  &  1.57 \% & 41.63 \% & 15.50 \% & 29 \%\\
\hline 
4 $\sigma$ $<$ CI $\leq$ 6 $\sigma$ &  5.02 \% & 26.47 \% & 14.79 \%& \multirow{2}{*}{71 \%}  \\
\cline{1-4}
 CI $>$ 6 $\sigma$ &  93.41 \% & 31.91 \%& 69.71 \% & \\
\hline\hline
\end{tabular}
\end{center}\caption{The summary of prediction efficiency for different methods. 
The initial MCMC samples are collected based on $\chi_{\rm{scan}}^2$ 
but re-weighted by using $\chi^2_{\rm{tot}}$ where $\chi_{\rm{scan}}^2=\chi^2_{\rm{tot}}/10$. 
The network predictions of {\bf Machine A} and {\bf Machine B} are based on 
the pseudo-data created by using $\chi^2_{\rm{tot}}$. 
For a comparison, we build a classifier ``Check Machine" by combining total collected mock data. }
\label{tab:Data_Summary_v2}
\end{table}

Finally, we compare the prediction efficiency for different methods
if the chi-square is updated to the latest experimental value, then $\chi_{\rm{scan}}^2\to \chi^2_{\rm{tot}}$.
In Table~\ref{tab:Data_Summary_v2}, it is not surprising that 
only a minor group $6.5\%$ of total MCMC samples can survive from $6\sigma$ cut of $\chi^2_{\rm{tot}}$,
because the initial MCMC samples are created by using a loose condition $\chi_{\rm{scan}}^2=\chi^2_{\rm{tot}}/10$.
Remarkably, {\bf Machine A} achieves good performance in that approximately $70\%$ 
of the total predictions fall into the $6\sigma$ region 
while this percentage is approximately $30\%$ for {\bf Machine B}.
Even though we train both machines by using MCMC data,
it is still much more efficient to generate samples in a new allowed region than simply reweighting the old MCMC data.

As an additional comparison, we build a typical classifier ``Check Machine" to identify
whether the model parameters fall into the $4\sigma$ region that is determined by using all the data samples collected in this study.
The details of the ``Check Machine" are given in Appendix~\ref{sec:checkmachine}.
We note that the Check Machine can identify the $4\sigma$ parameter space better than {\bf Machine B}.
Evidently, {\bf Machine A} is still the most efficient network to predict the parameter space that falls into the required CI.

%%%%%%%%%%%%%%%%%%%%%%%%%%%%%%%%%%%%%
\section{Summary and discussion}\label{sec:conclusion}
%%%%%%%%%%%%%%%%%%%%%%%%%%%%%%%%%%%%%

In this paper, we investigated the inverse function of the
CR diffusion equation by using a CNN network.
This function allowed us to efficiently and directly determine CR source functions and propagation mechanisms.
Subsequently, we created two CNN machines to obtain the inverse function of the CR diffusion equation.
In general, {\bf Machine A} learned the mock data directly, 
while {\bf Machine B} learned the mock data with added artificial fluctuations.
The systematic uncertainties of the two machines introduced by the
learning procedures were evaluated. 
We found that the procedure of identifying artificial fluctuations may
somewhat disturb the inverse propagation in {\bf Machine B}.

We prepared the learning data of the two CNN machines by using an MCMC global fitting based on
the measured spectra from the secondary cosmic ray nuclei Li, Be, B, C, and O from AMS-02, ACE, and Voyager-1.
However, we wanted to demonstrate that our networks can be used even for future updated data.
Therefore, scans were performed with the total chi-square divided by 10, and the result was used as learning data.
These two machines were adjusted as much as possible to obtain the best performance.

If the pseudospectra fed into the networks were created based on a new dataset with a different $\chi^2$,
two machines could reasonably predict the CR model parameters at the required confidence interval.
When data were updated, we only required a $4\sigma$ confidence interval as the best demonstration of our initial MCMC data.
By giving a new experimental dataset with smaller error bars,
both {\bf Machine A} and {\bf Machine B} performed much more efficiently in finding the required confidence interval
in the multidimensional parameter space than the one obtained by reweighting old MCMC data.
In addition, we also compared the two machines with a typical classifier that can identify the $4\sigma$ region
in the multidimensional CR parameter space by learning the reweighted old MCMC data.
We found that our {\bf Machine A} still performed better than this classifier, while {\bf Machine B} performs slightly poorer.

In summary, the greatest advantage of our method is that
when a new dataset is introduced or an existing dataset is updated,
we do not need to retrain the networks nor repeat the whole MCMC analysis.
The trained model can easily propose an approximate likelihood distribution of the CR model parameters. 
Furthermore, to demonstrate the usefulness of the inverse function of the propagation equation in this work,
we have only fed pseudospectra into the network.
However, if future detectors precisely measure the CR spectra for the whole energy range, 
the experimental spectra can be directly used for the network inputs. 
In addition, other spectra, such as anti-proton, Helium-3 and Helium-4 can be included in the training, 
but currently, these spectra suffer from large uncertainty~\cite{Wu:2018lqu}.

Here, we comment on our choice of the initial mock data, where
the chi-square metrics of the old data are designed to be ten times larger than the one being updated.
This assumption is indeed conservative compared to the actual situation 
because future CR experiments might only improve the error bars by lesser factor.
Under such an update, our pseudospectra shall be more reliable so that
our networks can provide confidence interval with more controls.

Although we only focus on CR propagation in this paper, our approach can be 
easily adapted to solve the inverse problem of multidimensional parameter 
space in other fields, for example, searching for supersymmetry at the LHC~\cite{ArkaniHamed:2005px}.
Moreover, as soon as a standard CR propagation model is built, 
our machines can be simply adapted for searching for the signature of dark matter 
from precisely measured cosmic ray experimental data.
Finally, we suggest three interesting applications of using our machines: 
\begin{itemize}
\item Our machines can be generators of the model parameters with 
  a proposed prior probability distribution. This usage is shown in this paper.
  \item To find the correlations between any physical spectra and theoretical model parameters,
  the architecture in Fig.~\ref{Fig:Architecture} can be reused for new training.
  This can save much time in tuning the layer structure and hyperparameters.
  \item Based on the latest experimental measurements,
  our inverse function networks of CR propagation can suggest suitable CR model parameters for a nonstandard source search,
  such as DM or any new cosmic ray source.
The machines allow us to put the latest experimental constraints on the DM parameter space by computing only the DM contribution.
\end{itemize}

 \subsection*{Acknowledgments}
We would like to thank Chi Chan for helpful discussions at the early
stages of this work. We thank Roberto Ruiz de Austri Bazan, Sai-Ping
Li and Wei-Chih Huang for useful comments and suggestions.
Y.-L.~S.~Tsai was funded in part by the Taiwan Young Talent Programme
of Chinese Academy of Sciences under Grant No.~2018TW2JA0005 and the
Ministry of Science and Technology, Taiwan under Grant
No.~109-2112-M-007-022-MY3. Q.Y. is supported by the NSFC under Grants
No. 11722328, No. 11851305, the 100 Talents program of Chinese Academy
of Sciences, and the Program for Innovative Talents and Entrepreneur
in Jiangsu. The work of K.C. was supported by the National Science
Council of Taiwan under Grant No.~MOST-107-2112-M-007-029-MY3. 
\newpage
\appendix

\section{Supplemental figures}
\label{appendix:total_phase_plane}

\subsection{The kinetic energy distribution}
\label{app:CRflux}

\setcounter{subfigure}{0}
\begin{figure}[ht]
\centering
\subfigure[Lithium]{
  \includegraphics[width=0.4\textwidth] {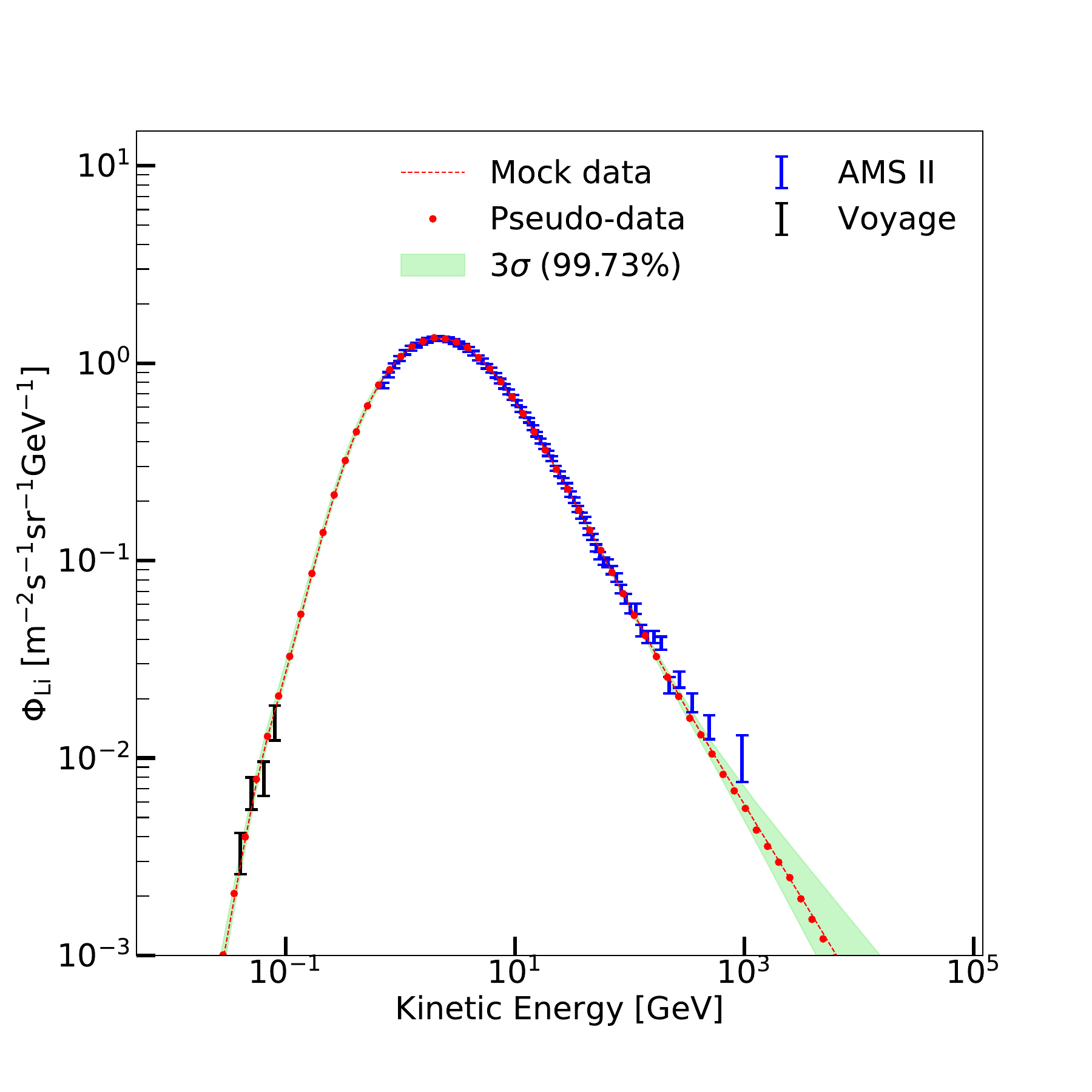}
 }
 \subfigure[Beryllium]{
  \includegraphics[width=0.4\textwidth] {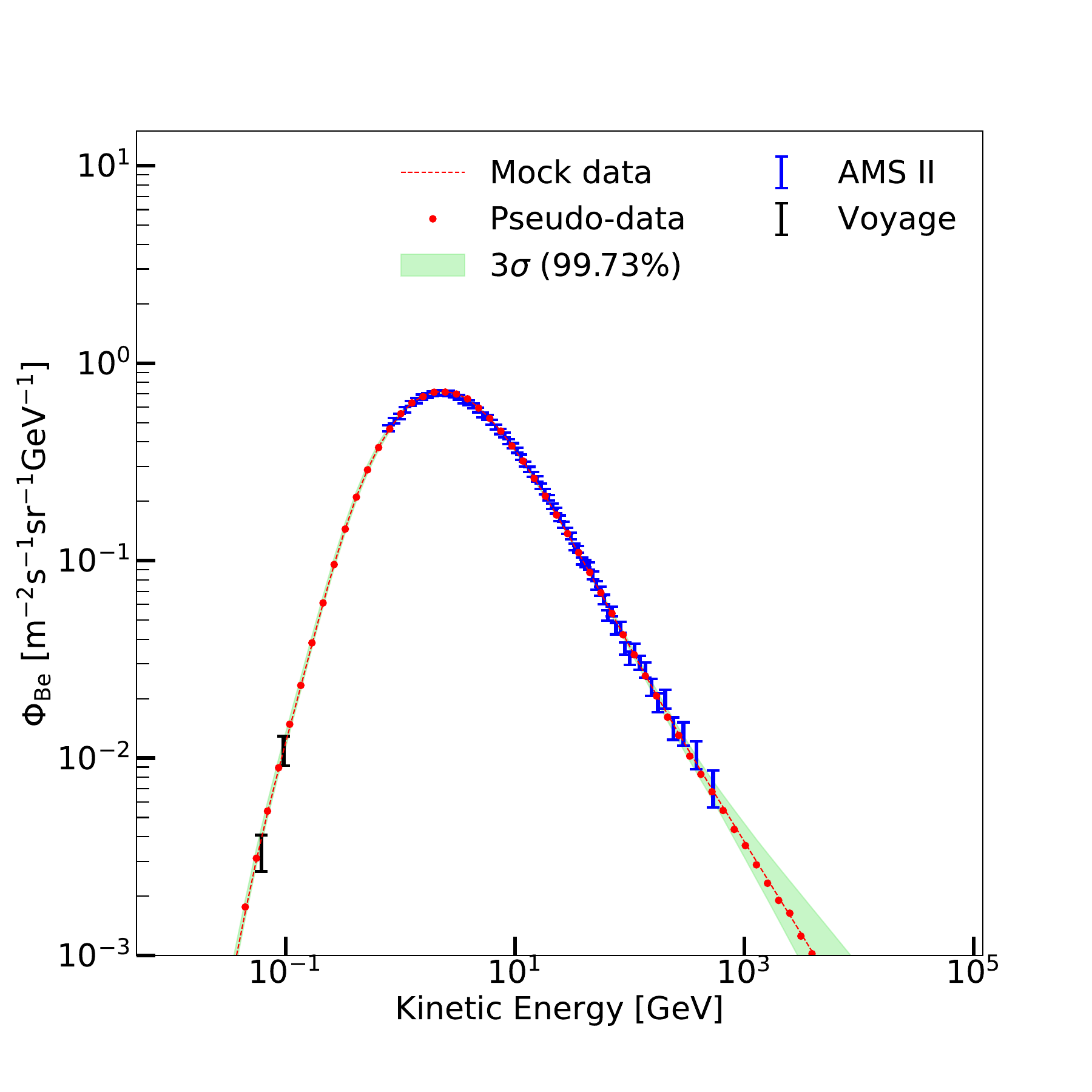}
 }
  \subfigure[Boron]{
  \includegraphics[width=0.4\textwidth] {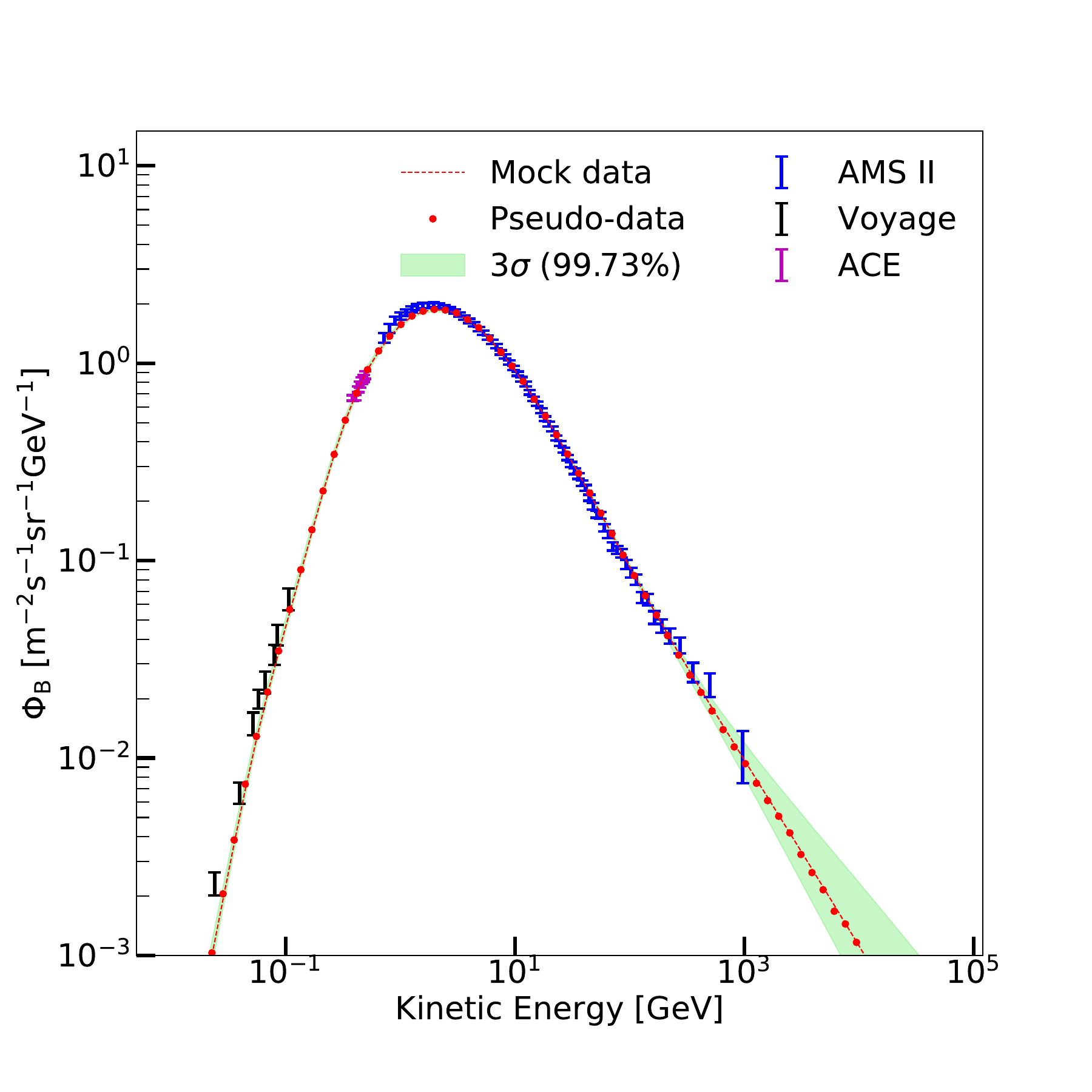}
 }
  \subfigure[Oxygen]{
  \includegraphics[width=0.4\textwidth] {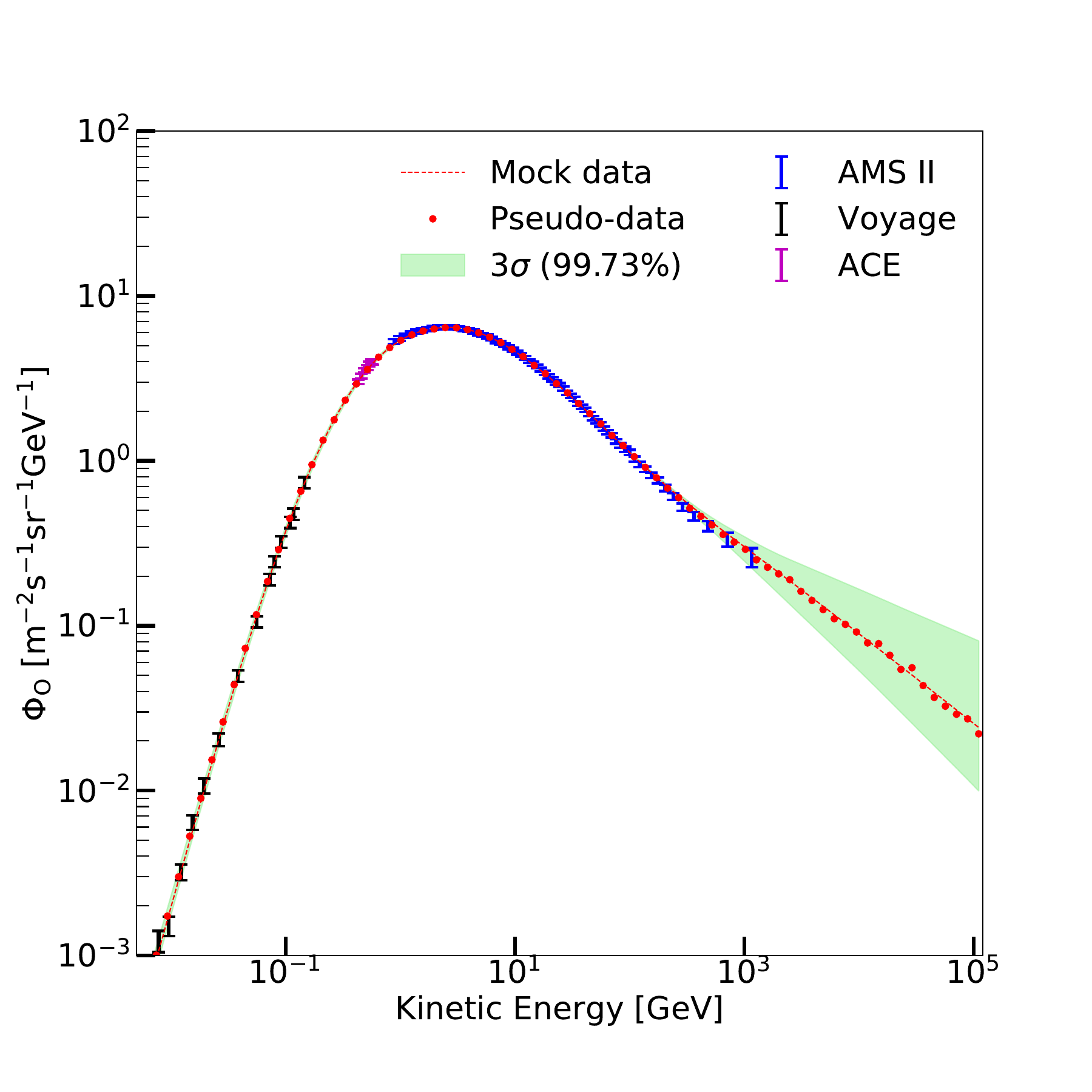}
 }
\caption{\label{fig:appCRflux}
 The demodulated kinetic energy spectra of Li, Be, B, and O.
 The solar modulation parameter is fixed at $\phi=696~{\rm MV}$. }
\end{figure}

In Fig.~\ref{fig:appCRflux}, the demodulated fluxes when $\phi=696~{\rm MV}$ for four different
elements, Li (upper left), Be (upper right), B (lower left), and O (lower right), are shown.
The green band indicates the $3\sigma$ uncertainty obtained
by the global analysis of the latest experimental data from AMS02 (blue), Voyage (black), and ACE (purple).
The $84$ bins of pseudo-spectrum (red dots) are evenly located at
the kinetic energy range $10^{-3}\gev<\log E<1.1\times 10^{5}\gev$.

\subsection{One- and two-dimensional likelihood distribution}
\label{sec:1d2dlike}

\begin{figure}[hb!]
\begin{center}
\includegraphics[width=0.32\textwidth]{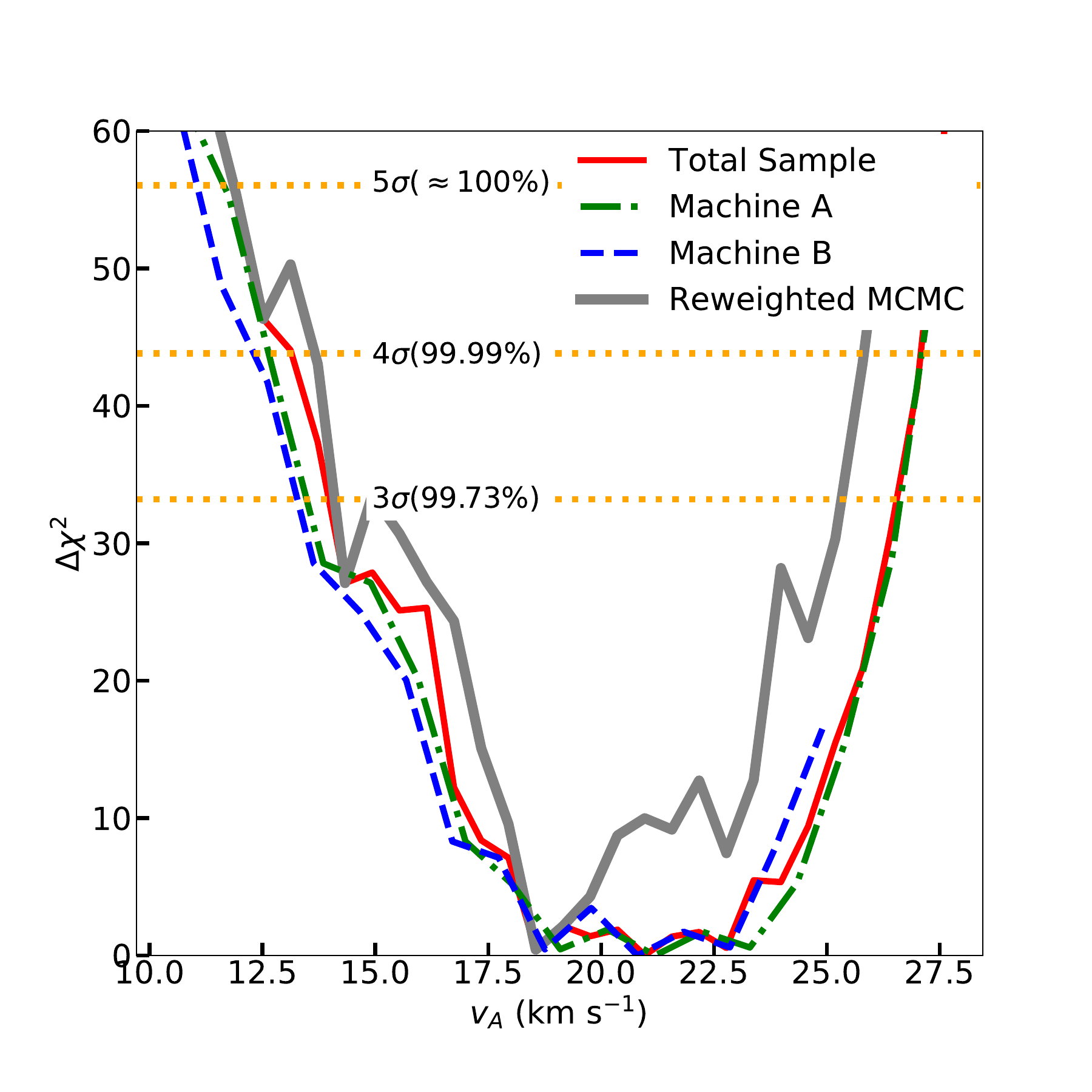}
\includegraphics[width=0.32\textwidth]{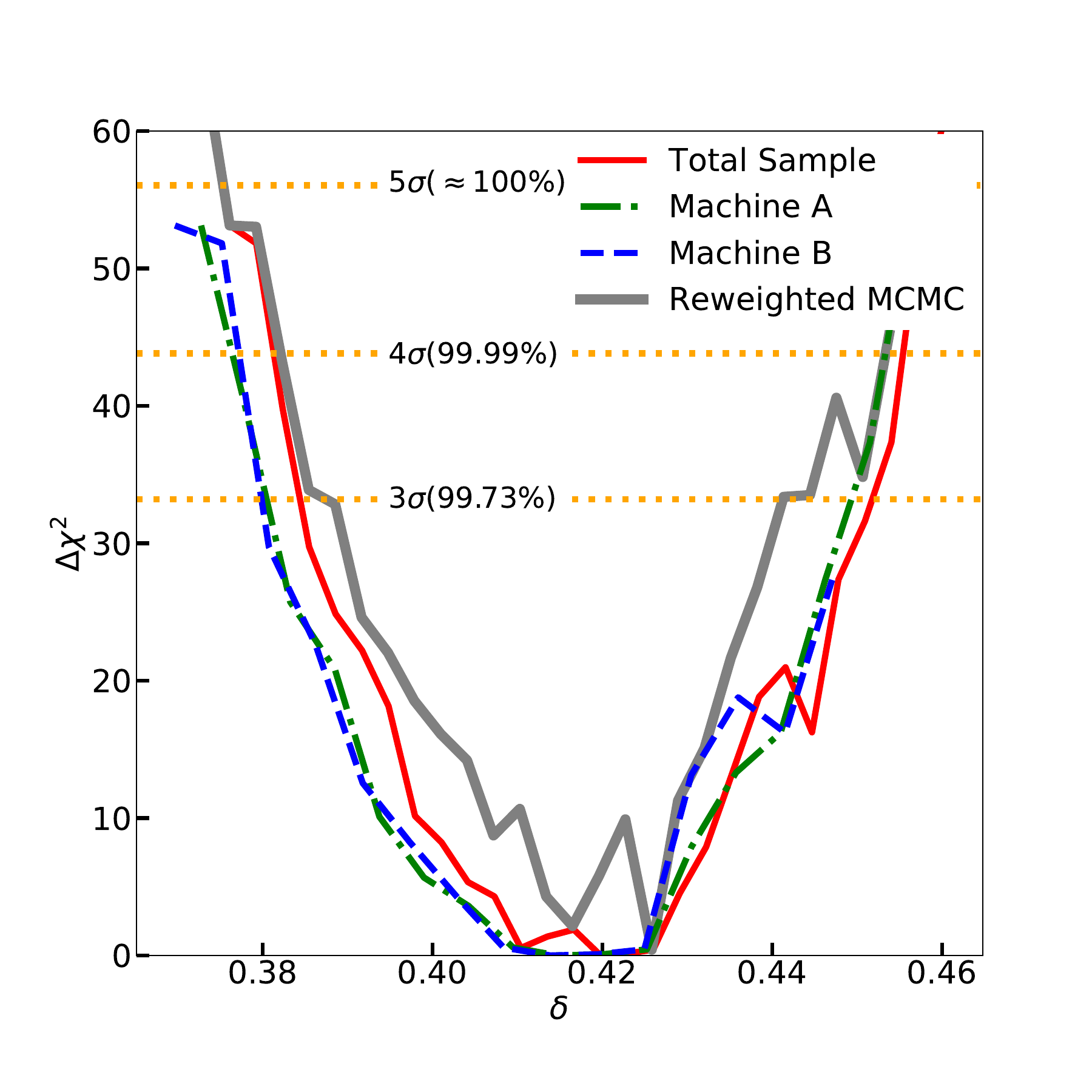}
\includegraphics[width=0.32\textwidth]{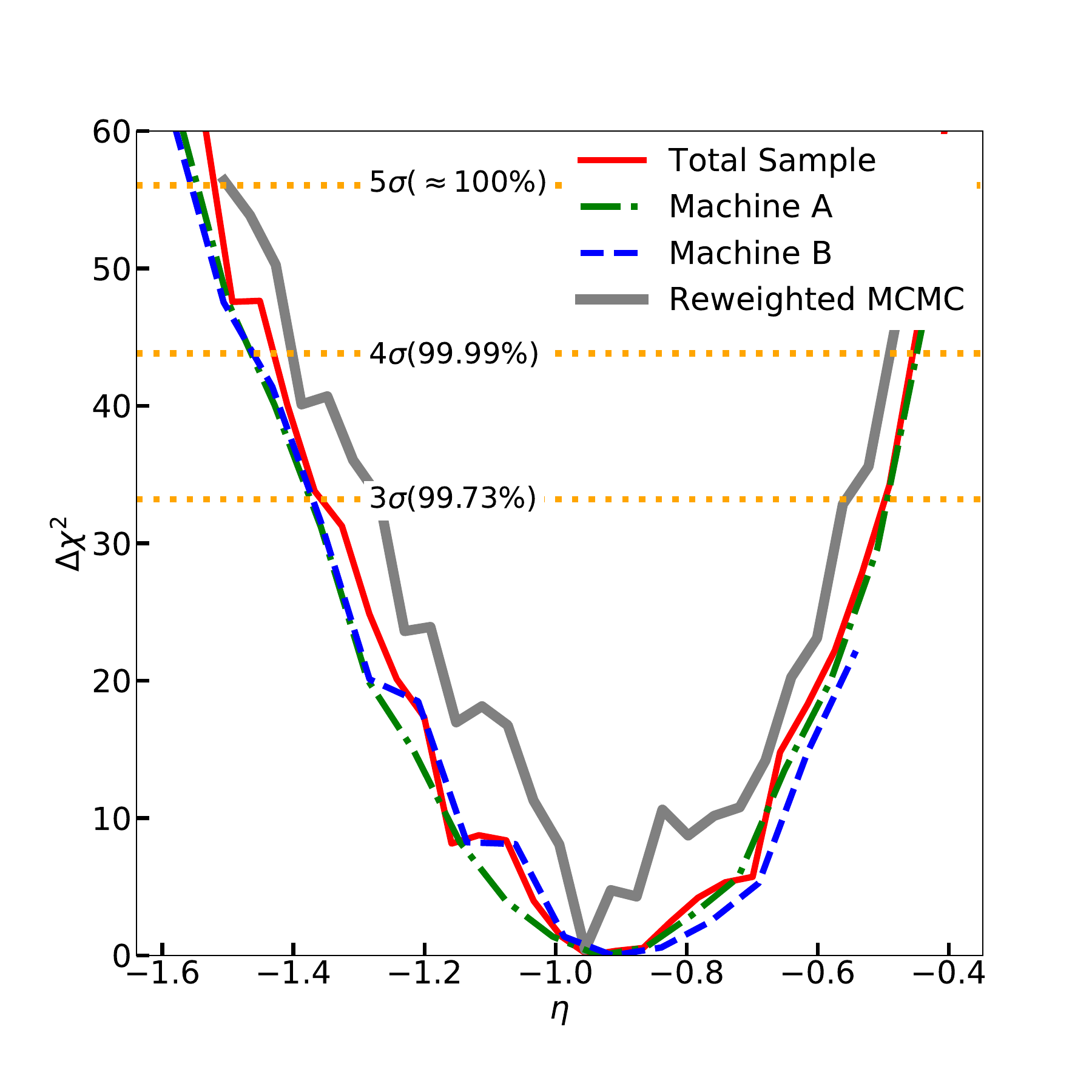}\\
\includegraphics[width=0.32\textwidth]{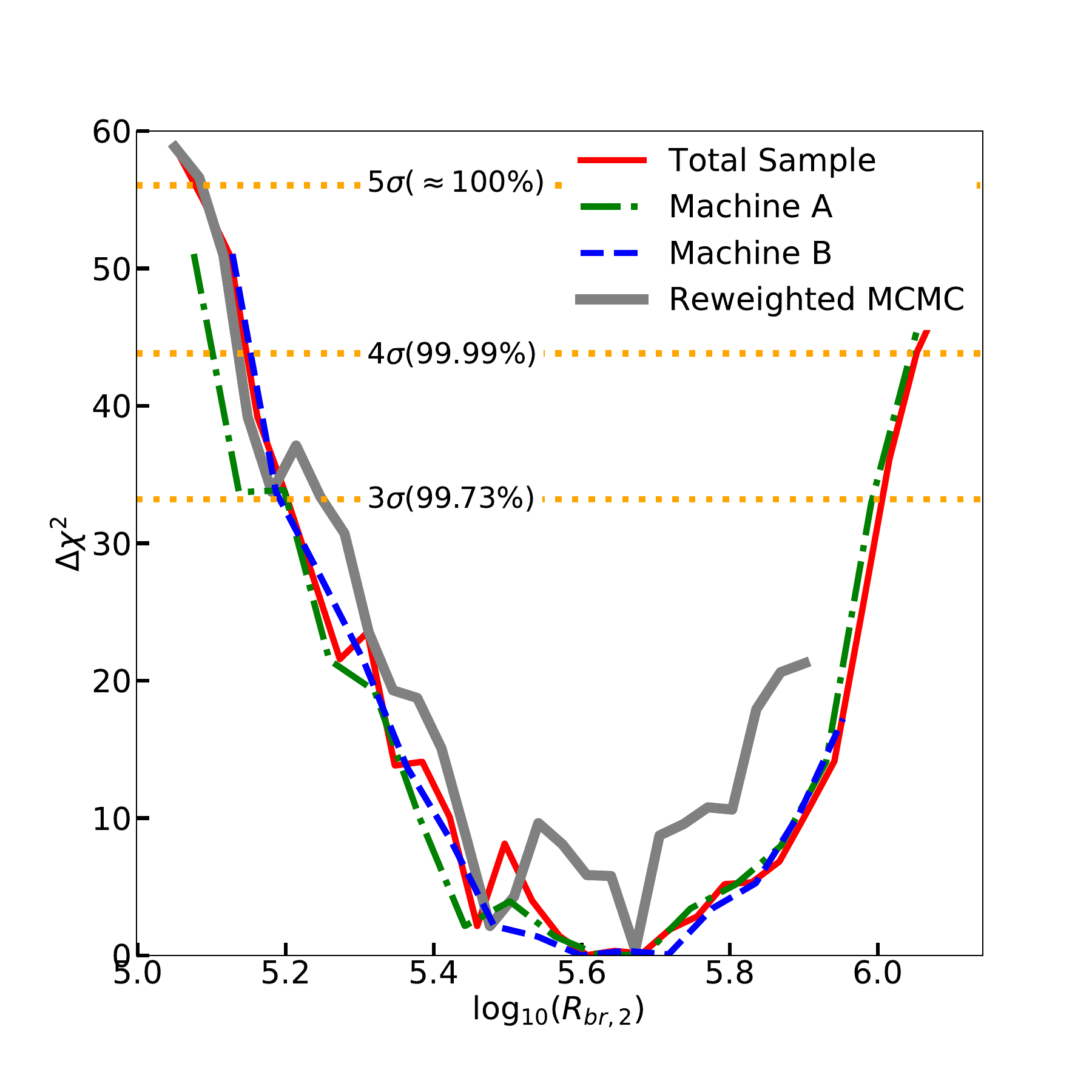} 
\includegraphics[width=0.32\textwidth]{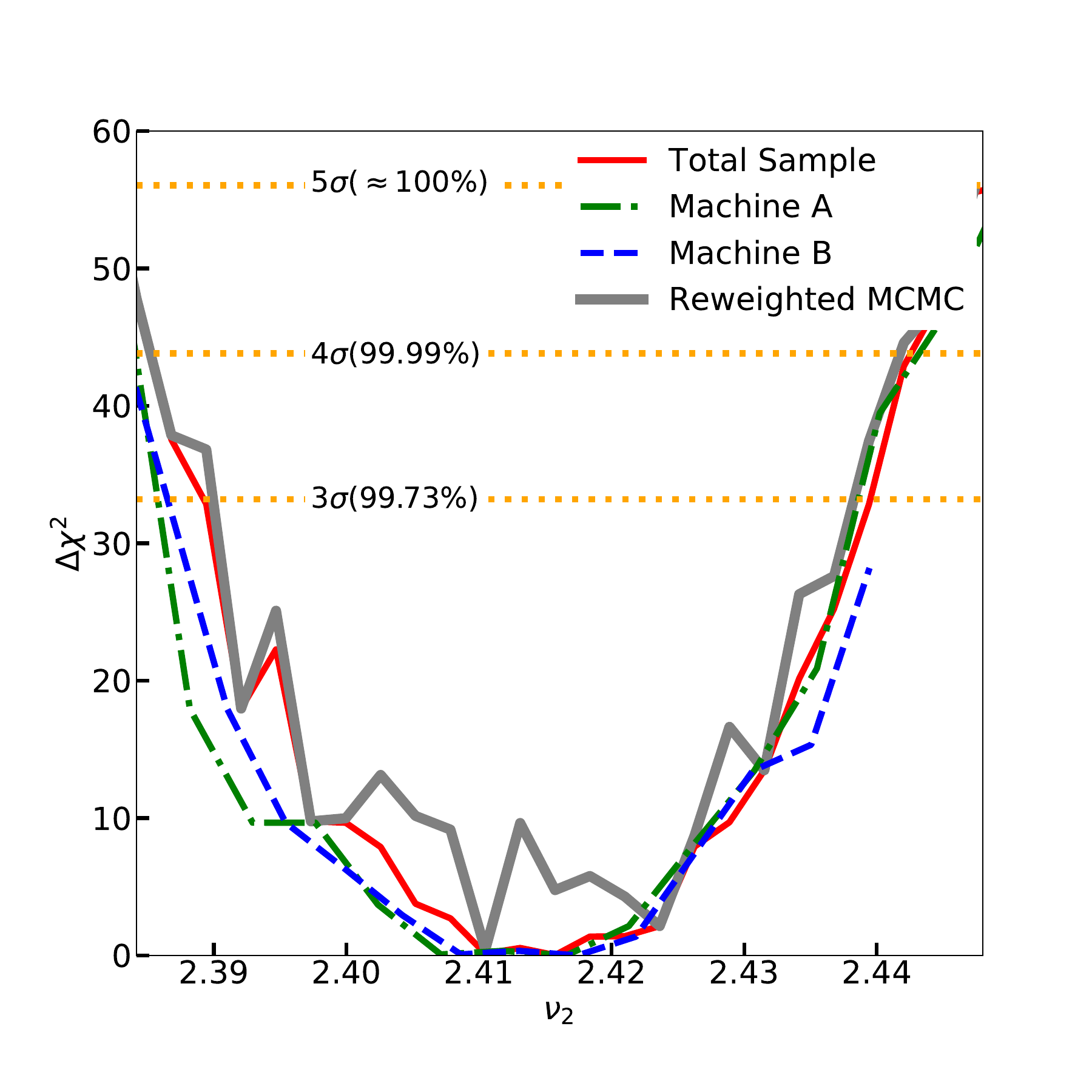}
\includegraphics[width=0.32\textwidth]{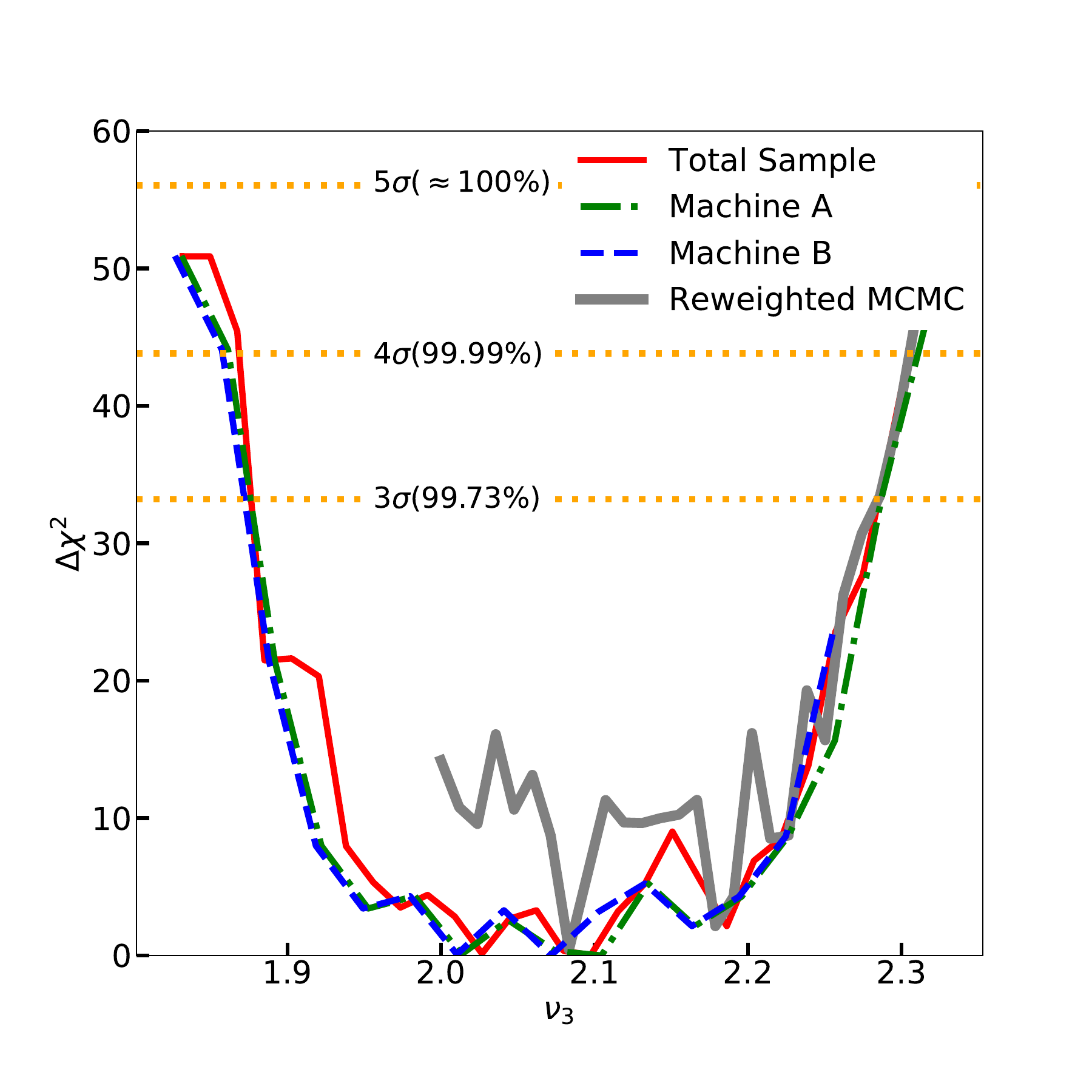} 
 \caption{\label{fig:1Dlike2}
The one-dimensional $\Delta\chi^2$ distributions for CR model parameters, 
except for $D_0$, $Z_h$, $\nu_1$, and $R_{br,1}$, which are shown in Fig.~\ref{fig:1Dlike}.
The grey thick solid, red solid, blue dashed, and green dash-dotted curves show the $\Delta\chi^2$ distribution
of the initial MCMC scan, all samples, the predictions of {\bf Machine A} and the predictions of {\bf Machine B}, respectively.
The CIs for $3\sigma$, $4\sigma$, and $5\sigma$ are shown by the orange horizontal dotted curves.
}
 \end{center}
 \end{figure}

In Fig.~\ref{fig:1Dlike2}, we present the one-dimensional $\Delta\chi^2$ distributions
for the six remaining CR model parameters (Eq.~\eqref{eq:modelpar}) in addition to Fig.~\ref{fig:1Dlike}.
The colour coding is the same as before.
The network predictions from both {\bf Machine A} and {\bf Machine B} mimic
the $\Delta\chi^2$ distributions very well.
Again, despite the wavy feature of MCMC caused by scan coverage,
our networks already capture the CR model features and predict much smoother curves.

\begin{figure}[hb!]
\begin{center}
\includegraphics[width=0.98\textwidth]{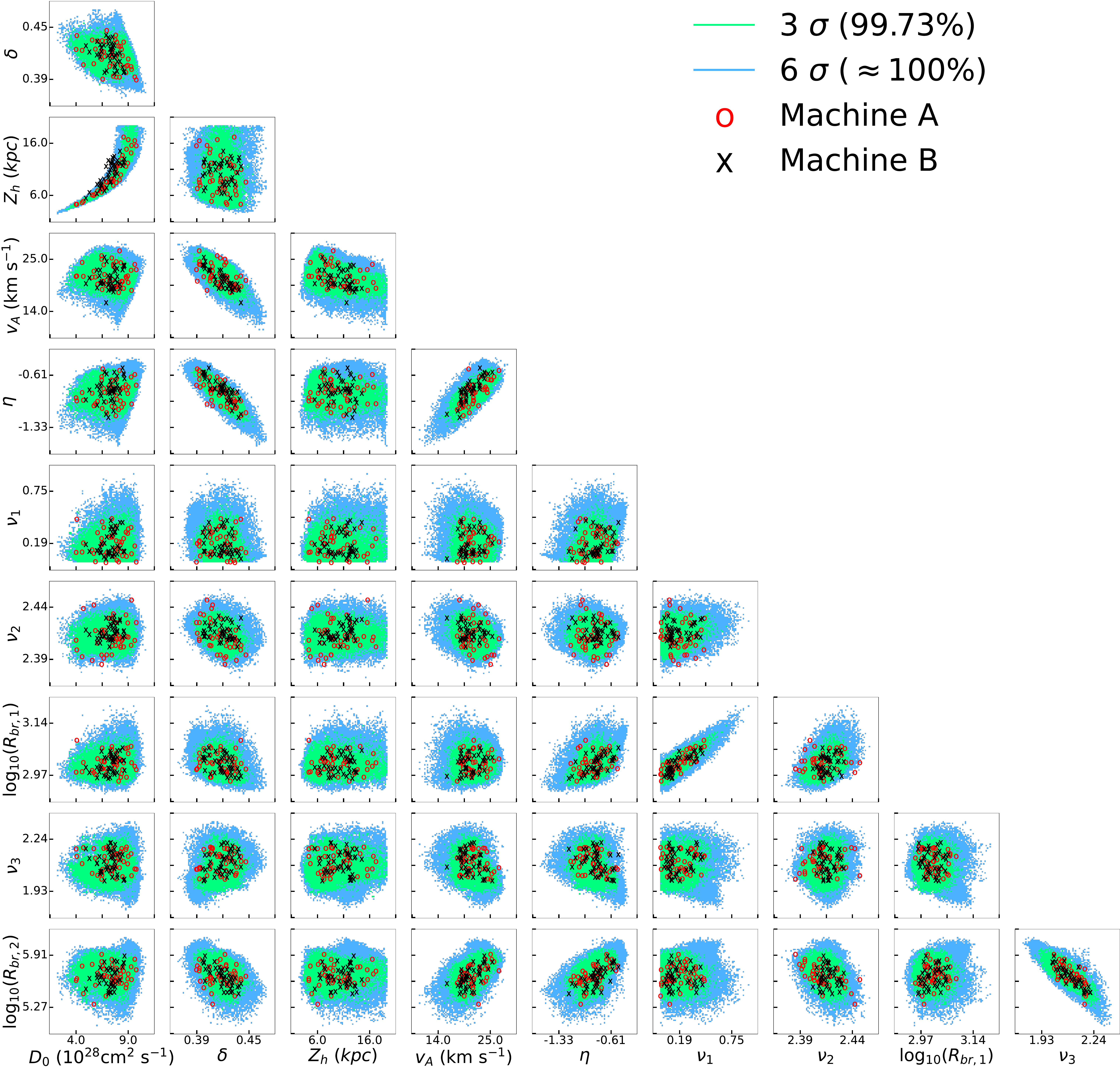}
\caption{Two-dimensional distributions of the propagation and source parameters.
The colour scheme is the same as in Fig.~\ref{Fig:phasespace}.
\label{Fig:total_phase_plane}}
\end{center}
\end{figure}

In Fig.~\ref{Fig:total_phase_plane}, we show the distribution of all the propagation and source parameters.
Similar to the colour scheme in Fig.~\ref{Fig:phasespace}, the green and blue regions represent the $3\sigma$
and $6\sigma$ CI of the latest cosmic ray data, respectively.
As a comparison, we also show $100$ predicted $4\sigma$ points by {\bf Machine A} (red circles)
and {\bf Machine B} (black crosses).
Clearly, the majority of predicted $4\sigma$ points ($3\sigma$ mock data plus $1\sigma$ artificial fluctuations)
agrees with the $3\sigma$ contours.
Although these predicted points are sampled on the experimental fluxes with respect to energies,
our machines still know the correlation between parameters.
Moreover, even the farthest points are still included within the $6\sigma$ CI.

%%%%%%%%%%%%%%%%%%%%
\section{Check Machine}
\label{sec:checkmachine}
%%%%%%%%%%%%%%%%%%%%

\begin{figure}[h]
\begin{center}
\includegraphics[width=0.5\textwidth]{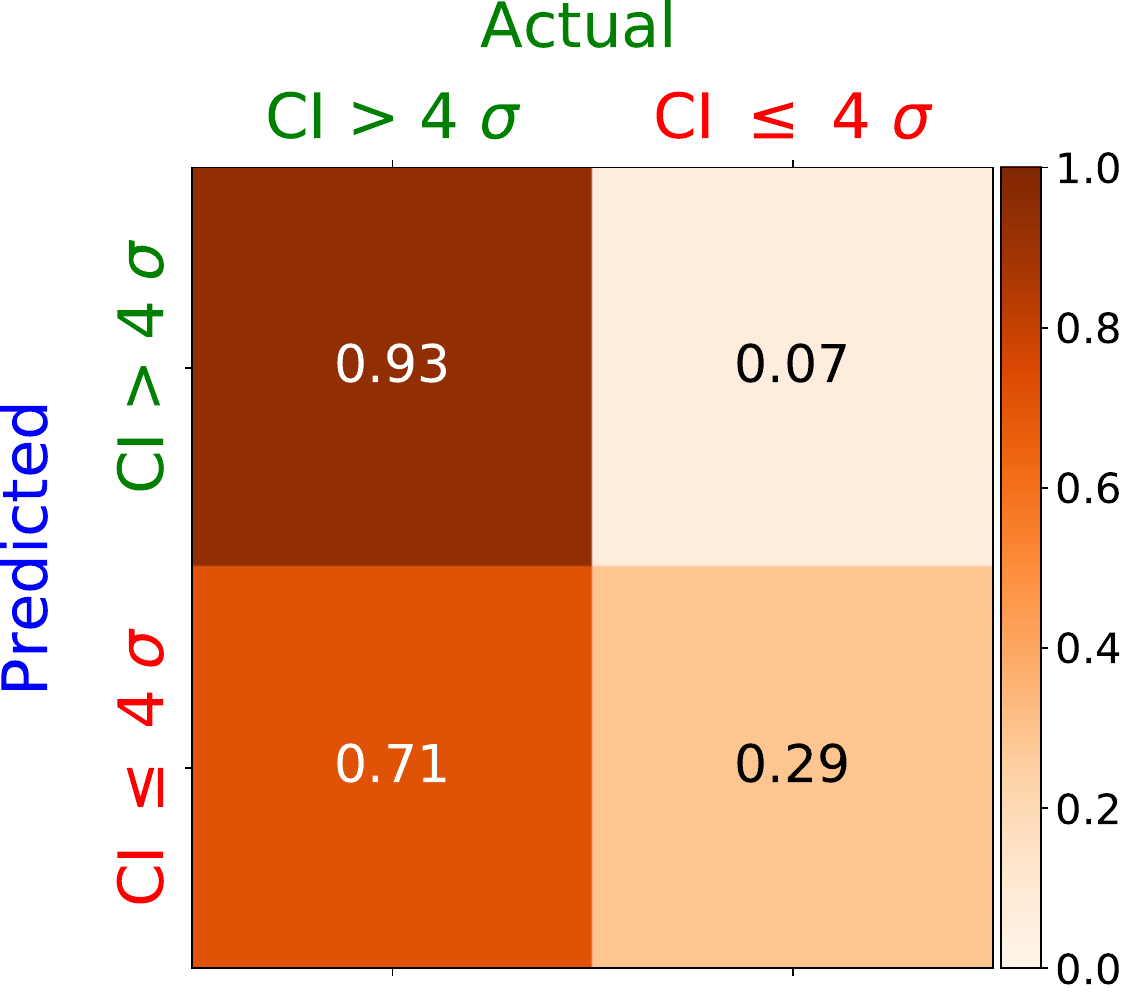}
\caption{The confusion matrix of the Check Machine.
The Check Machine is trained with source and propagation parameters
to find parameter sets within the $4\sigma$ region.
We obtain a $29\%$ true positive.
\label{Fig:confusion_matrix}}
\end{center}
\end{figure}

This machine aims to determine whether the parameter set $\{\theta_i\}$ is within the $4\sigma$ region
directly from the source and propagation parameter space.
For the training data,
we select those samples within the $4\sigma$ region and
randomly select an equal number of samples outside the $4\sigma$ region
from the total collected mock data.
Unlike {\bf Machine A} and {\bf Machine B},
we use the source and propagation parameters as the input data of training.
The data are divided into $90\%$ and $10\%$ for training and validation, respectively.
The Check Machine contains six fully connected dense layers with the \texttt{ReLU} activation function,
However, we use binary cross entropy as the loss function for binary class purposes.
After adjusting the performance,
we found that only $29\%$ of the true data (inside the $4\sigma$ region) can be identified
by the Check Machine, while $93\%$ of the data outside the $4\sigma$ region can be
identified as false.
We show the confusion matrix in Fig.~\ref{Fig:confusion_matrix}.


\newpage
\begin{thebibliography}{15} 

\bibitem{Stone:2013}
  E.~C. Stone, A.~C. Cummings, F.~B. McDonald, B.~C. Heikkila, N. Lal, W.~R. Webber,
  %``Voyager 1 Observes Low-Energy Galactic Cosmic Rays in a Region Depleted of Heliospheric Ions,''
  Science \textbf{341}, 150 (2013)
  %doi:10.1126/science.1236408

%\cite{Maurin:2001sj}
\bibitem{Maurin:2001sj} 
  D.~Maurin, F.~Donato, R.~Taillet and P.~Salati,
  %``Cosmic rays below z=30 in a diffusion model: new constraints on propagation parameters,''
  Astrophys.\ J.\  {\bf 555}, 585 (2001)
  %doi:10.1086/321496
  [astro-ph/0101231].
  %%CITATION = doi:10.1086/321496;%%
  %373 citations counted in INSPIRE as of 27 Aug 2020

%\cite{Trotta:2010mx}
\bibitem{Trotta:2010mx}
  R.~Trotta, G.~Jóhannesson, I.~V.~Moskalenko, T.~A.~Porter, R.~R.~d.~Austri and A.~W.~Strong,
  %``Constraints on cosmic-ray propagation models from a global Bayesian analysis,''
  Astrophys. J. \textbf{729}, 106 (2011)
  %doi:10.1088/0004-637X/729/2/106
  [arXiv:1011.0037 [astro-ph.HE]].
  %229 citations counted in INSPIRE as of 03 Sep 2020

%\cite{Yuan:2017ozr}
\bibitem{Yuan:2017ozr} 
  Q.~Yuan, S.~J.~Lin, K.~Fang and X.~J.~Bi,
  %``Propagation of cosmic rays in the AMS-02 era,''
  Phys.\ Rev.\ D {\bf 95}, no. 8, 083007 (2017)
  %doi:10.1103/PhysRevD.95.083007
  [arXiv:1701.06149 [astro-ph.HE]].
  %%CITATION = doi:10.1103/PhysRevD.95.083007;%%
  %52 citations counted in INSPIRE as of 27 Aug 2020
  
%\cite{Strong:1998pw}
\bibitem{Strong:1998pw} 
  A.~W.~Strong and I.~V.~Moskalenko,
  %``Propagation of cosmic-ray nucleons in the galaxy,''
  Astrophys.\ J.\  {\bf 509}, 212 (1998)
  %doi:10.1086/306470
  [astro-ph/9807150].
  %%CITATION = doi:10.1086/306470;%%
  %779 citations counted in INSPIRE as of 27 Aug 2020


%\cite{Evoli:2008dv}
\bibitem{Evoli:2008dv} 
  C.~Evoli, D.~Gaggero, D.~Grasso and L.~Maccione,
  %``Cosmic-Ray Nuclei, Antiprotons and Gamma-rays in the Galaxy: a New Diffusion Model,''
  JCAP {\bf 0810}, 018 (2008)
  Erratum: [JCAP {\bf 1604}, E01 (2016)]
  %doi:10.1088/1475-7516/2008/10/018, 10.1088/1475-7516/2016/04/E01
  [arXiv:0807.4730 [astro-ph]].
  %%CITATION = doi:10.1088/1475-7516/2008/10/018, 10.1088/1475-7516/2016/04/E01;%%
  %226 citations counted in INSPIRE as of 27 Aug 2020

%\cite{Cui:2016ppb}
\bibitem{Cui:2016ppb}
  M.~Y.~Cui, Q.~Yuan, Y.~L.~S.~Tsai and Y.~Z.~Fan,
  %``Possible dark matter annihilation signal in the AMS-02 antiproton data,''
  Phys. Rev. Lett. \textbf{118}, no.19, 191101 (2017)
  %doi:10.1103/PhysRevLett.118.191101
  [arXiv:1610.03840 [astro-ph.HE]].
  %113 citations counted in INSPIRE as of 03 Sep 2020

%\cite{Cuoco:2016eej}
  \bibitem{Cuoco:2016eej}
  A.~Cuoco, M.~Krämer and M.~Korsmeier,
  %``Novel Dark Matter Constraints from Antiprotons in Light of AMS-02,''
  Phys. Rev. Lett. \textbf{118}, no.19, 191102 (2017)
  %doi:10.1103/PhysRevLett.118.191102
  [arXiv:1610.03071 [astro-ph.HE]].
  %121 citations counted in INSPIRE as of 03 Sep 2020


%\cite{Lin:2019ljc}
\bibitem{Lin:2019ljc} 
  S.~J.~Lin, X.~J.~Bi and P.~F.~Yin,
  %``Investigating the dark matter signal in the cosmic ray antiproton flux with the machine learning method,''
  Phys.\ Rev.\ D {\bf 100}, no. 10, 103014 (2019)
  %doi:10.1103/PhysRevD.100.103014
  [arXiv:1903.09545 [astro-ph.HE]].
  %%CITATION = doi:10.1103/PhysRevD.100.103014;%%
  %5 citations counted in INSPIRE as of 27 Aug 2020

\bibitem{BOSER} 
  BOSER, B.E., GUYON, I.M. and VAPNIK, V.N. (1992). A Training Algorithm for Optimal Margin Classifiers. In {\it{Proceedings of the 5th Annual ACM Workshop on Computational Learning Theory}} 144-152. ACM Press.
  
\bibitem{VAPNIK} 
    CORTES, C. and VAPNIK, V. (1995). Support-Vector Networks. In {\it{Machine Learning}} 273-297.

\bibitem{Friedman} 
  J. H. Friedman, {\it{Stochastic gradient boosting, Computational Statistics \& Data Analysis}} 38 (2002), no. 4 367 - 378. Nonlinear Methods and Data Mining

  
\bibitem{Ridgeway} 
  G. Ridgeway, {\it{Generalized boosted models: A guide to the gbm package}}, 2006.

\bibitem{Chen_2016} 
    T. Chen and C. Guestrin, {\it{Xgboost, Proceedings of the 22nd ACM SIGKDD International Conference on Knowledge Discovery and Data Mining}} (Aug, 2016)

%\cite{Guest:2018yhq}
\bibitem{Guest:2018yhq}
D.~Guest, K.~Cranmer and D.~Whiteson,
%``Deep Learning and its Application to LHC Physics,''
Ann. Rev. Nucl. Part. Sci. \textbf{68}, 161-181 (2018)
doi:10.1146/annurev-nucl-101917-021019
[arXiv:1806.11484 [hep-ex]].
%84 citations counted in INSPIRE as of 02 Feb 2021

\bibitem{DBLP:journals/corr/abs-1812-08434}
    Jie Zhou, Ganqu Cui, Zhengyan Zhang, Cheng Yang, Zhiyuan Liu, Lifeng Wang, Changcheng Li, Maosong Sun,
    [arXiv:1812.08434 [cs.LG]].
    

\bibitem{DBLP:journals/corr/abs-1808-03314}
    Alex Sherstinsky
    [arXiv:1808.03314 [cs.LG]].

%\cite{Fenton:2020woz}
\bibitem{Polosukhin}
    Polosukhin, Illia; Kaiser, Lukasz; Gomez, Aidan N.; Jones, Llion; Uszkoreit, Jakob; Parmar, Niki; Shazeer, Noam; Vaswani, Ashish,
    [arXiv:1706.03762 [cs.CL]].

%\cite{Fenton:2020woz}
\bibitem{Fenton:2020woz}
M.~J.~Fenton, A.~Shmakov, T.~W.~Ho, S.~C.~Hsu, D.~Whiteson and P.~Baldi,
%``Permutationless Many-Jet Event Reconstruction with Symmetry Preserving Attention Networks,''
[arXiv:2010.09206 [hep-ex]].
%2 citations counted in INSPIRE as of 30 Jan 2021

%\cite{Ren:2017ymm}
\bibitem{Ren:2017ymm} 
  J.~Ren, L.~Wu, J.~M.~Yang and J.~Zhao,
  %``Exploring supersymmetry with machine learning,''
  Nucl.\ Phys.\ B {\bf 943}, 114613 (2019)
  %doi:10.1016/j.nuclphysb.2019.114613
  [arXiv:1708.06615 [hep-ph]].
  %%CITATION = doi:10.1016/j.nuclphysb.2019.114613;%%
  %20 citations counted in INSPIRE as of 27 Aug 2020


%\cite{Abdughani:2019wuv}
\bibitem{Abdughani:2019wuv} 
  M.~Abdughani, J.~Ren, L.~Wu, J.~M.~Yang and J.~Zhao,
  %``Supervised deep learning in high energy phenomenology: a mini review,''
  Commun.\ Theor.\ Phys.\  {\bf 71}, no. 8, 955 (2019)
  %doi:10.1088/0253-6102/71/8/955
  [arXiv:1905.06047 [hep-ph]].
  %%CITATION = doi:10.1088/0253-6102/71/8/955;%%
  %14 citations counted in INSPIRE as of 27 Aug 2020


%\cite{Alsing:2019xrx}
\bibitem{Alsing:2019xrx} 
  J.~Alsing, T.~Charnock, S.~Feeney and B.~Wandelt,
  %``Fast likelihood-free cosmology with neural density estimators and active learning,''
  Mon.\ Not.\ Roy.\ Astron.\ Soc.\  {\bf 488}, no. 3, 4440 (2019)
  %doi:10.1093/mnras/stz1960
  [arXiv:1903.00007 [astro-ph.CO]].
  %%CITATION = doi:10.1093/mnras/stz1960;%%
  %15 citations counted in INSPIRE as of 27 Aug 2020


%\cite{Lei:2020ucb}
\bibitem{Lei:2020ucb} 
  Y.~K.~Lei, C.~Liu and Z.~Chen,
  %``Numerical analysis of neutrino physics within a high scale supersymmetry model via machine learning,''
  arXiv:2006.01495 [hep-ph].
  %%CITATION = ARXIV:2006.01495;%%

\bibitem{Yuan:2018vgk}
Q.~Yuan,
%``Implications on cosmic ray injection and propagation parameters from Voyager/ACE/AMS-02 nucleus data,''
Sci. China Phys. Mech. Astron. \textbf{62}, no.4, 49511 (2019)
%doi:10.1007/s11433-018-9300-0
[arXiv:1805.10649 [astro-ph.HE]].

%\cite{Strong:2007nh}
\bibitem{Strong:2007nh} 
  A.~W.~Strong, I.~V.~Moskalenko and V.~S.~Ptuskin,
  %``Cosmic-ray propagation and interactions in the Galaxy,''
  Ann.\ Rev.\ Nucl.\ Part.\ Sci.\  {\bf 57}, 285 (2007)
  %doi:10.1146/annurev.nucl.57.090506.123011
  [astro-ph/0701517].
  %%CITATION = doi:10.1146/annurev.nucl.57.090506.123011;%%
  %745 citations counted in INSPIRE as of 27 Aug 2020


%\cite{Maurin:2010zp}
\bibitem{Maurin:2010zp} 
  D.~Maurin, A.~Putze and L.~Derome,
  %``Systematic uncertainties on the cosmic-ray transport parameters: Is it possible to reconcile B/C data with delta = 1/3 or delta = 1/2?,''
  Astron.\ Astrophys.\  {\bf 516}, A67 (2010)
  %doi:10.1051/0004-6361/201014011
  [arXiv:1001.0553 [astro-ph.HE]].
  %%CITATION = doi:10.1051/0004-6361/201014011;%%
  %74 citations counted in INSPIRE as of 27 Aug 2020

\bibitem{DiBernardo:2009ku}
G.~Di Bernardo, C.~Evoli, D.~Gaggero, D.~Grasso and L.~Maccione,
%``Unified interpretation of cosmic-ray nuclei and antiproton recent measurements,''
Astropart. Phys. \textbf{34}, 274-283 (2010)
%doi:10.1016/j.astropartphys.2010.08.006
[arXiv:0909.4548 [astro-ph.HE]].

\bibitem{Seo:1994}
E.~S.~Seo, V.~S.~Ptuskin, Astrophys.\ J.\  {\bf 431}, 705 (1994)

%\cite{Yuan:2018lmc}
\bibitem{Yuan:2018lmc}
Q.~Yuan, C.~R.~Zhu, X.~J.~Bi and D.~M.~Wei,
%``Secondary cosmic-ray nucleus spectra disfavor particle transport in the Galaxy without reacceleration,''
JCAP \textbf{11} (2020), 027
doi:10.1088/1475-7516/2020/11/027
[arXiv:1810.03141 [astro-ph.HE]].
%12 citations counted in INSPIRE as of 24 Jan 2022


%\cite{Putze:2010zn}
\bibitem{Putze:2010zn} 
  A.~Putze, L.~Derome and D.~Maurin,
  %``A Markov Chain Monte Carlo technique to sample transport and source parameters of Galactic cosmic rays: II. Results for the diffusion model combining B/C and radioactive nuclei,''
  Astron.\ Astrophys.\  {\bf 516}, A66 (2010)
  %doi:10.1051/0004-6361/201014010
  [arXiv:1001.0551 [astro-ph.HE]].
  %%CITATION = doi:10.1051/0004-6361/201014010;%%
  %114 citations counted in INSPIRE as of 27 Aug 2020

\bibitem{Aguilar:2017hno}
M.~Aguilar \textit{et al.} [AMS],
%``Observation of the Identical Rigidity Dependence of He, C, and O Cosmic Rays at High Rigidities by the Alpha Magnetic Spectrometer on the International Space Station,''
Phys. Rev. Lett. \textbf{119}, no.25, 251101 (2017)
%doi:10.1103/PhysRevLett.119.251101

\bibitem{Aguilar:2018njt}
M.~Aguilar \textit{et al.} [AMS],
%``Observation of New Properties of Secondary Cosmic Rays Lithium, Beryllium, and Boron by the Alpha Magnetic Spectrometer on the International Space Station,''
Phys. Rev. Lett. \textbf{120}, no.2, 021101 (2018)
%doi:10.1103/PhysRevLett.120.021101

\bibitem{ACE}
N.~E.~Yanasak et al., Astrophys.\ J.\  {\bf 563}, 768 (2001)

\bibitem{Zhu:2018jbk}
C.~R.~Zhu, Q.~Yuan and D.~M.~Wei,
%``Studies on cosmic ray nuclei with Voyager, ACE and AMS-02: I. local interstellar spectra and solar modulation,''
Astrophys. J. \textbf{863}, no.2, 119 (2018)
%doi:10.3847/1538-4357/aacff9
[arXiv:1807.09470 [astro-ph.HE]].

\bibitem{Cummings:2016pdr}
A.~C.~Cummings, E.~C.~Stone, B.~C.~Heikkila, N.~Lal, W.~R.~Webber, G.~Jóhannesson, I.~V.~Moskalenko, E.~Orlando and T.~A.~Porter,
%``Galactic Cosmic Rays in the Local Interstellar Medium: Voyager 1 Observations and Model Results,''
Astrophys. J. \textbf{831}, no.1, 18 (2016)
%doi:10.3847/0004-637X/831/1/18

%\cite{maxpooling}
\bibitem{maxpooling}
Kouichi Yamaguchi, Kenji Sakamoto, Toshio Akabane, Yoshiji Fujimoto, (1990): A neural network for speaker-independent isolated word recognition, In ICSLP-1990, 1077-1080.

\bibitem{Liu:2011re}
J.~Liu, Q.~Yuan, X.~J.~Bi, H.~Li and X.~Zhang,
%``CosRayMC: a global fitting method in studying the properties of the new sources of cosmic e$^{\pm}$ excesses,''
Phys. Rev. D \textbf{85}, 043507 (2012)
%doi:10.1103/PhysRevD.85.043507
[arXiv:1106.3882 [astro-ph.CO]].



  
%\cite{ELU}
\bibitem{ELU} 
  Daeho Kim, Jinah Kim, Jaeil Kim,
  Elastic exponential linear units for convolutional neural networks, Neurocomputing, Volume 406,2020


%\cite{Adam}
\bibitem{Adam} 
  Diederik P. Kingma and Jimmy Lei Ba,
  ADAM: A METHOD FOR STOCHASTIC OPTIMIZATION,
  [arXiv:1412.6980v9 [cs.LG]].



%\cite{tensorflow2015-whitepaper}
\bibitem{tensorflow2015-whitepaper} 
    Martín Abadi, Ashish Agarwal, Paul Barham, Eugene Brevdo,
    Zhifeng Chen, Craig Citro, Greg S. Corrado, Andy Davis,
    Jeffrey Dean, Matthieu Devin, Sanjay Ghemawat, Ian Goodfellow,
    Andrew Harp, Geoffrey Irving, Michael Isard, Rafal Jozefowicz, Yangqing Jia,
    Lukasz Kaiser, Manjunath Kudlur, Josh Levenberg, Dan Mané, Mike Schuster,
    Rajat Monga, Sherry Moore, Derek Murray, Chris Olah, Jonathon Shlens,
    Benoit Steiner, Ilya Sutskever, Kunal Talwar, Paul Tucker,
    Vincent Vanhoucke, Vijay Vasudevan, Fernanda Viégas,
    Oriol Vinyals, Pete Warden, Martin Wattenberg, Martin Wicke,
    Yuan Yu, and Xiaoqiang Zheng.
    TensorFlow: Large-scale machine learning on heterogeneous systems,
    2015. Software available from tensorflow.org.
  

%\cite{Goodfellow:2014upx}
\bibitem{Goodfellow:2014upx}
I.~J.~Goodfellow, J.~Pouget-Abadie, M.~Mirza, B.~Xu, D.~Warde-Farley, S.~Ozair, A.~Courville and Y.~Bengio,
%``Generative Adversarial Networks,''
[arXiv:1406.2661 [stat.ML]].
%91 citations counted in INSPIRE as of 03 Sep 2020

%\cite{Butter:2019cae}
\bibitem{Butter:2019cae}
A.~Butter, T.~Plehn and R.~Winterhalder,
%``How to GAN LHC Events,''
SciPost Phys. \textbf{7}, no.6, 075 (2019)
doi:10.21468/SciPostPhys.7.6.075
[arXiv:1907.03764 [hep-ph]].
%29 citations counted in INSPIRE as of 12 Oct 2020


%\cite{Wu:2018lqu}
\bibitem{Wu:2018lqu}
J.~Wu and H.~Chen,
%``Revisit cosmic ray propagation by using $^{1}$H, $^{2}$H, $^{3}$He and $^{4}$He,''
Phys. Lett. B \textbf{789}, 292-299 (2019)
doi:10.1016/j.physletb.2018.11.052
[arXiv:1809.04905 [astro-ph.HE]].
%6 citations counted in INSPIRE as of 15 Feb 2021



%\cite{ArkaniHamed:2005px}
\bibitem{ArkaniHamed:2005px}
N.~Arkani-Hamed, G.~L.~Kane, J.~Thaler and L.~T.~Wang,
%``Supersymmetry and the LHC inverse problem,''
JHEP \textbf{08} (2006), 070
%doi:10.1088/1126-6708/2006/08/070
[arXiv:hep-ph/0512190 [hep-ph]].
%133 citations counted in INSPIRE as of 02 Sep 2020
  

\end{thebibliography}
\end{document}